\documentclass{aa}  

\usepackage{amsmath}
\usepackage{amssymb}
\usepackage{appendix}
\usepackage{nicefrac}
\usepackage{multicol}
\usepackage{xcolor}
\usepackage{graphicx}
\usepackage{txfonts}

\DeclareMathOperator\erf{erf}
\DeclareMathOperator\erfi{erfi}
\newcommand{\iu}{{i\mkern1mu}}

\begin{document} 

\title{The impact of the point spread function  fitting radius on photometric uncertainty based on the Fisher information matrix}

\author{Sebastian Espinosa\inst{1,2}, Mario L.   Vicuña\inst{2}, Rene A. Mendez\inst{3}, Jorge F. Silva\inst{2} \and Marcos Orchard\inst{2}}

\institute{Advanced center for Electrical and Electronic Engineering, Universidad Técnica Federico Santa María, Gral. Bari 699, Valparaiso, Chile \\
\email{sebastian.espinosat@usm.cl}
\and
Departamento de Ingeniería Eléctrica, Universidad de Chile, Av. Tupper, Santiago, Chile\\
\email{josilva@ing.uchile.cl}
\and
Departamento de Astronomia, Universidad de Chile. Casilla 36-D, Santiago, Chile\\
\email{rmendez@uchile.cl}
}

\abstract
{In point spread function (PSF) photometry, the selection of the fitting aperture radius plays a critical role in determining the precision of flux and background estimations. Traditional methods often rely on maximizing the signal-to-noise ratio (S/N) as a criterion for aperture selection. However, S/N-based approaches do not necessarily provide the optimal precision for joint estimation problems as they do not account for the statistical limits imposed by the Fisher information in the context of the Cramér-Rao lower bound (CRLB).}
{This study aims to establish an alternative criterion for selecting the optimal fitting radius based on Fisher information rather than S/N. Fisher information serves as a fundamental measure of estimation precision, providing theoretical guarantees on the achievable accuracy for parameter estimation. By leveraging Fisher information, we seek to define an aperture selection strategy that minimizes the loss of precision.}
{We conducted a series of numerical experiments that analyze the behavior of Fisher information and estimator performance as a function of the PSF aperture radius. Specifically, we revisited fundamental photometric models and explored the relationship between aperture size and information content. We compared the empirical variance of classical estimators, such as maximum likelihood and stochastic weighted least squares, against the theoretical CRLB derived from the Fisher information matrix.}
{Our results indicate that aperture selection based on the Fisher information provides a more robust framework for achieving optimal estimation precision. The findings reveal that S/N-based aperture selection may lead to significant discrepancies, with potential precision losses of up to 70\%. In contrast, Fisher information-based selection allows   a more accurate and consistent estimation process, ensuring that the empirical variance closely aligns with the theoretical limits.}
 {}

\titlerunning{Impact of  PSF fitting radius on photometric uncertainty based on the Fisher information matrix}
\authorrunning{Espinosa et al.} 
\maketitle
\section{Introduction} \label{sec:intro}

Photometry is an important technique in astrophysics used to measure the intensity or brightness of celestial objects, such as solar system bodies, stars, and galaxies \citep{1992ASSL..175.....S, 1995SSRv...73..437E, budding2007introduction}. 
In recent decades, the advent of space missions dedicated to high-precision photometric monitoring, such as \emph{CoRoT} \citep{auvergne2009corot}, \emph{Kepler} \citep{borucki2010kepler}, its extended \emph{K2} mission \citep{howell2014k2}, and \emph{TESS} \citep{ricker2015transiting},  have transformed the field by enabling unprecedented photometric precision and continuity in time series. 
These capabilities are critical for applications such as detecting and characterizing exoplanets, probing stellar oscillations, and studying long-term stellar variability. 
In the near future, the \emph{PLATO} mission \citep{rauer2014plato} and, farther ahead, the proposed \emph{HAYDN} project \citep{2021ExA....51..963M},\footnote{See an up-to-date description of the project at \newline \url{https://www.asterochronometry.eu/haydn/index.html}} aim to extend this legacy by delivering photometry with even higher precision, longer baselines, and improved characterization of systematic effects. 
Such missions require highly optimized photometric extraction methods, as their precision goals approach the fundamental limits imposed by photon statistics and the stability of the instrument.

Photometric measurements are essential for various fields of study, from detecting and characterizing exoplanets via transits to monitoring variable stars and supernovae; these measurements can help unravel stellar evolutions or cosmological distances \citep{gallaway2020introduction, kitchin2020astrophysical}. As light from celestial sources often gets contaminated by background noise, whether from the sky, adjacent objects, or instrumental interference, accurate photometry requires techniques that can effectively isolate the target's light from this background \citep{2010tms..book.....C,2011ASSL..373....1S}. Traditional methods perform these estimations in a sequential two-step manner \citep{1999ASPC..189...50M}, first estimating the background and then the source flux (see, e.g., DAOPHOT in \citet{1987PASP...99..191S}, DOPHOT in \citet{1993PASP..105.1342S}, MATPHOT in \citet{2005MNRAS.361..861M}, or PSFEx in \citet{2011ASPC..442..435B} using a parametric approach, and \citet{2011ASPC..442..107C} for a Bayesian approach, among others). This two-step approach has been demonstrated by \citet{2024PASP..136a4501V} to introduce biases and limit the attainable precision, proposing instead a novel methodology for estimating background and source flux simultaneously for aperture photometry, applicable to isolated sources. For this paper we adopted that methodology, but we now address the issue of the impact in photometric precision of the choice of aperture when performing point spread function (PSF) fitting in crowded fields.

Classical photometric techniques have long relied on the optimization of the signal-to-noise ratio (S/N) as a key determinant of accuracy, particularly when using charged-coupled devices (CCD) for point-like source observations. Traditional methods, such as those outlined by \citet{1989PASP..101..616H,1991PASP..103..122N,1998MNRAS.296..339N}, have made significant contributions to improving photometric precision by carefully managing noise sources, including sky background, dark current, and readout noise. Howell’s introduction of the CCD equation (see also \citet{1995ExA.....6..163M}, their Eq.~(21)), which accounts for various noise terms beyond simple Poisson statistics, and Newberry’s thorough treatment of noise propagation in sky-subtracted data have provided astronomers with valuable tools for improving photometric precision. These methods often involve using a small aperture to deal with crowded fields, where aperture photometry is unfeasible, while maximizing the S/N. As we shall demonstrate below, this approach is not always a good benchmark indicator, and can severely limit photometry precision, particularly when there is some flexibility in choosing the PSF fitting radius.

Modern photometry methods, based on the Cramér-Rao lower bound (CRLB), have been employed as a key indicator of photometric precision \cite{1983AJ.....88.1683L,adorf1996limits}. More importantly, estimators have been proposed that achieve this bound with extremely high accuracy, where the precision is a direct function of the number of observations and, consequently, the aperture radius. This represents a paradigm shift in photometric techniques as it indicates that to achieve the highest possible precision \citep{2013PASP..125..580M,2014PASP..126..798M,2018A&A...616A..95E} it is necessary to utilize the maximum amount of information available, which implies using the entire usable image around the target for photometric analysis---if possible---rather than limiting the data to pre-selected pixels or regions. This approach challenges traditional methods and underscores the importance of accessing and incorporating all available relevant pixel data to refine flux and background estimations, leading to more precise and accurate results in astronomical photometry. For example, \cite{perryman1989improved} suggested that the precision of flux measurements is improved by better accounting for both the background and the source's flux simultaneously, leading to more accurate and efficient photometric pipelines.

In \citet{2024PASP..136a4501V}, they show that the classical approach of determining first the background in an area surrounding the source and subtracting that estimation from the source to quantify the flux of the target is sub-optimal in comparison with a joint simultaneous estimation of these quantities. This was demonstrated from theoretical grounds and based on extensive numerical simulations under realistic observational scenarios. They  also tested their methodology on very high-precision photometry based on TESS satellite data on isolated sources, confirming their theoretical predictions. To our knowledge, this is the first completely self-consistent treatment of this problem, on solid statistical grounds, which gives us theoretical assurances of optimality (in the CRLB sense) as well as unbiasedness on the derived quantities of interest, namely, source flux and background.

Given the results presented in the previous paragraph, it becomes essential to examine how the selection of the aperture radius during PSF fitting influences the final photometric precision. In many practical scenarios, such as crowded fields or varying seeing conditions, the choice of aperture radius cannot be arbitrary and often plays a pivotal role in balancing precision and feasibility. Despite its importance, this aspect is commonly guided by heuristics, which may not always reflect the true limits of photometric accuracy.

In this paper we propose a novel methodology that reframes the problem of aperture selection from an information-theoretic perspective. Instead of relying on S/N, we compute the CRLB for the joint estimation of source flux and background level, thereby quantifying the theoretical precision limit associated with each aperture choice. The Fisher information, which underpins the CRLB, provides a principled measure of the data's informativeness and allows  a rigorous assessment of the trade-offs involved in aperture selection. The main contributions of this work are summarized below.

 \begin{itemize} 
\item We introduce an information-theoretic approach to photometry based on the Fisher information matrix. This approach represents a fundamental paradigm shift in how we approach photometry, moving beyond S/N to a more robust statistical foundation that enables higher precision and flexibility in observational strategies while, at the same time, offering guarantees of precision and unbiased results in a quantifiable way.

\item By basing our approach on this statistical framework, we are able to quantify in a precise manner how the choice of a given PSF fitting radius impacts on the accuracy of flux measurements.
\item We demonstrate that the proposed method consistently outperforms traditional S/N-based approaches, especially for larger apertures where bias and underestimation of uncertainty are common. 
\item Our results provide a data-driven tool to inform PSF aperture selection, offering both improved accuracy and flexibility for a wide range of observational constraints. \end{itemize}

The paper is organized as follows. In Sect.~\ref{sec:preliminaries} we revisit the fundamental results of the astrometric model for characterizing point sources on a CCD. In Sect.~\ref{sec:aperture} we formalize the optimal aperture analysis and present the main results. In Sect.~\ref{numerical}  we show numerical results analyzing discrepancies between the performance based on Fisher information and some classical practical estimators. Finally, in Sect.~\ref{sec:conclusions} we offer our conclusions and final remarks on our findings.

\section{Preliminaries} \label{sec:preliminaries}

We present a summary of the results obtained by \citet{2024PASP..136a4501V} in the context of the joint source and background flux. For completeness, the subsequent section will describe the model in some detail, with a focus on emphasizing the most relevant results that will be critical for the following section.

\subsection{Statistical parametric model for photometry} \label{sub_sec_astro_photo}

The primary focus is the inference of the flux and background of a point source. In our simplified but realistic model, the source is characterized by two scalar parameters: the object's position $x_c\in \mathbb{R}$ within the array, and its aggregated intensity,\footnote{Also usually named as brightness or flux, accumulated on a certain exposure time. To be more precise, in this treatment it refers to the total photo-e$^{-}$ from the source, in counts. See Table~\ref{tab:baseline}.} denoted as $F\in \mathbb{R}^+$. These two parameters define a probability distribution $\mu_{x_c,F}$ over an observation space, denoted as $\mathbb{X}$. Formally, a point source specified by the pair $(x_c,F)$ generates a nominal intensity profile on a photon-integrating device (PID), such as a CCD, which can be expressed as 
\begin{equation}\label{eq_pre_1}
F_{x_c}(x)=F \cdot \phi(x-x_c,\sigma),
\end{equation}
where $\phi(x-x_c,\sigma)$ represents the one-dimensional normalized point spread function (PSF), with $\sigma$ serving as a generic parameter that defines the width (or spread) of the light distribution on the detector. This parameter is generally influenced by the wavelength and the observing site's quality  \citep{2013PASP..125..580M,2014PASP..126..798M}.

The profile in Eq.~(\ref{eq_pre_1}) is not observable directly; rather, it is subject to the following three sources of perturbations:

\begin{itemize}
\item an additive background noise, which accounts for photon emissions from the open (diffuse) sky $f_s$, as well as noise introduced by the instrument itself, including read-out noise ($RON$) and dark current ($D$) (see, e.g., \citet{2001sccd.book.....J,2006hca..book.....H,2007ptd..book.....J,2008eiad.book.....M}), modeled by $B$ in
Eq.~(\ref{eq:background}):\footnote{This is also measured in e$^{-}$ counts. This model assumes that noise in each pixel is purely Poisson-distributed, which is appropriate when photon noise dominates. 
In practice, the readout noise introduced by the detector is well described by a Gaussian distribution. A more realistic noise model would therefore combine Poisson-distributed photon noise with Gaussian-distributed readout noise, resulting in a Poisson–Gaussian mixture and a correspondingly more complex likelihood function.
The Anscombe transform \citep{1948JRSS...Anscombe} and its extensions to mixed Poisson–Gaussian models \citep{1995Murtagh} 
offer practical approximations in such cases. 
In this work, we focus on the Poisson-limited case to maintain analytical tractability, leaving the full mixed-noise treatment for future work.} 
\begin{equation}
B = f_{s} \cdot \Delta x + D + RON^{2} \,; \label{eq:background}
\end{equation}

\item an uncertainty between the aggregated intensity (comprising the nominal object brightness and the background) and the actual measurements. Independent random variables model this uncertainty following a Poisson distribution (an analysis similar to the one that follows, but for Gaussian noise typical of photographic plates, was done by \citet{1983AJ.....88.1683L}, including the calculation of the Fisher matrix, see their Eq.~(8));

\item the spatial quantization process associated with the pixel resolution of the PID, as specified in Eqs.~(\ref{eq_pre_2b}) and (\ref{eq_pre_3}). 
\end{itemize}
By modeling these effects, we have a countable collection of independent and non-identically distributed random variables (observations or counts) $\left\{I_i: i \in
\mathbb{Z}\right\}$, where $I_i \sim Poisson(\lambda_i(F,B))$, driven by the expected intensity at each pixel element $i$,\footnote{For simplicity we assume here a linear detector, but the extension to two-dimensions is trivial, as it simply adds summations in two-coordinates. Alternatively, the two-dimensional case can be expressed with one index that encompasses the whole ordered two-dimensional array.} given by
\begin{equation}\label{eq_pre_2b}
\lambda_i(F,B) \equiv\mathbb{E}\{I_i\}= F \cdot g_i(x_c) + B_i,~\forall i\in \mathbb{Z},
\end{equation}
\begin{equation}\label{eq_pre_3}
g_i(x_c) \equiv \int^{x_i+\Delta x/2}_{x_i-\Delta x/2} \phi(x- x_c,\sigma)~d x, \ \forall i \in \mathbb{Z},
\end{equation}
\begin{equation}
\phi(x,\sigma) = \frac{1}{\sqrt{2\pi}\sigma}\exp{\left( \frac{-x^{2}}{2\sigma^{2}} \right)},\label{eq:gaussian_PSF}
\end{equation}
where $\mathbb{E}\left\lbrace \right\rbrace$ is the expectation value of the argument and $\left\{x_i: i \in \mathbb{Z}\right\}$ denotes the standard uniform quantization of the real line array with resolution $\Delta x>0$, i.e., $x_{i+1}-x_i=\Delta x$ for all $i \in \mathbb{Z}$. The function $g_i(x_c)$ in Eq. (\ref{eq_pre_3}) describes the distribution of (source) flux on the detector or, equivalently, the image profile across pixels, integrated over pixel $i$. 

In practice, the PID has a finite collection of measurement elements (or pixels). Then we  consider a finite set $\mathcal{N}~=~\{1,\dots,n\}$ of indices over the pixels (for simplicity, in what follows we  assume that $n$ is odd) of a CCD imaging a point source whose relative position $x_{c}$ is known, and $n~=~|\mathcal{N}|$. Another usual convention is to characterize the spread (width) of the PSF by its full width at half maximum ($FWHM$), which relates to the parameter $\sigma$ through
\begin{equation}
FWHM = 2 \sqrt{2\ln{2}} \sigma \,.\label{eq:FWHM} \end{equation}
Adopting this parametric model, the likelihood function of $\mathbf{I}=\{I_1,...,I_n\}$, given our target parameter vector $(F, B)$, is given by
\begin{equation}
L(\mathbf{I};F,B, \mathcal{N}) = \prod_{i\in\mathcal{N}} P_{\lambda_{i}(F,B)}(I_{i}) \, \label{eq:likelihood},
\end{equation}
\begin{equation}
P_{\lambda_{i}(F,B)}(I) = \frac{e^{-\lambda_{i}(F,B)} \cdot \lambda_{i}^{I}(F,B)}{I!},
\end{equation}
where $P_{\lambda}(x)=\frac{e^{-\lambda}\cdot \lambda^x}{x!}$ denotes
the probability mass function (PMF) of the Poisson law evaluated at the realization of $I=x$ (see, e.g.,  \cite{kay1993fundamentals}). Equation (\ref{eq:likelihood}) depends on the selected set of pixels. However, due to the independence of the observations, the likelihood will only change based on the observations chosen, not the order in which they are selected. In the following sections, we  consider subsets $\mathcal{J}_u \subset \mathcal{N}$ of pixels of the form $\mathcal{J}_u~=~\{n_{x_{c}} - n_{u}, \dots, n_{x_{c}}, \dots, n_{x_{c}} + n_{u}\}$, where $n_{u}$ is the number of full pixels within the aperture radius we have, and $n_{x_c}$ is the number of the pixel associated with the position of $x_c$ (typically $n/2$ if the array is centered).  Finally, $x_c$ is assumed to be known and the estimation task reduces to finding the parameter estimator $\hat{\boldsymbol{\theta}}~=~(\hat{F}, \hat{B})$, which is the inference of the underlying parameters $(F,B)$ from $\mathbf{I}$. Formally, the inference task consists in defining a regression rule $\tau(\mathbf{I}):~\mathbb{N}^{n}~\to~\boldsymbol{\Theta}~=~\mathbb{R}^{+} \times \mathbb{R}^{+}$, such that $(\hat{F}, \hat{B}) \equiv \tau(\mathbf{I})$.

\subsection{The Cramér-Rao lower bound (CRLB)} \label{sec:CRLB_theory}

The well-known CRLB offers a performance bound on the variance estimation error) of the family of unbiased estimators.\footnote{In the sense that $\mathbb{E}\{ \hat{\boldsymbol{\theta}} \} = \boldsymbol{\theta}$.} Here, we revisit the multidimensional\footnote{In the sense that various parameters are being determined.} version of this result.

\begin{theorem} \cite{rao1945, cramer1946} 
Let $\mathcal{N}$ be a finite set of integers with $|\mathcal{N}|=n$; $\{I_{i}\}_{i \in \mathcal{N}}$ be a collection of independent observations, whose likelihood function $L(\cdot; \boldsymbol{\theta},\mathcal{N})$ is induced by a parameter vector $\boldsymbol{\theta} = (\theta_{1},\dots,\theta_{m}) \in \boldsymbol{\Theta}, m \in \mathbb{N}$ over a parameter space $\boldsymbol{\Theta}$ (typically $\boldsymbol{\Theta} = \mathbb{R}^{m}$), such that the following regularity condition is satisfied:
\begin{equation}
\mathbb{E}\left\{ \frac{\partial \ln{L(\mathbf{I};\boldsymbol{\theta},\mathcal{N})}}{\partial \theta_{i}} \right\} = 0 \,, \hspace{1mm} \forall i \in \{1,\dots,m\} \,, \forall \boldsymbol{\theta} \in \boldsymbol{\Theta}. \label{eq:CRLB_reg}
\end{equation}
(i) Then, any unbiased estimator $\hat{\boldsymbol{\theta}}$ of $\boldsymbol{\theta}$, given by a regression rule $\tau: \mathbb{N}^{n} \to \boldsymbol{\Theta}, \hat{\boldsymbol{\theta}} \equiv \tau(\mathbf{I})$, has a covariance matrix $K_{\hat{\boldsymbol{\theta}}}\equiv\mathbb{E} \left\{ \left( \hat{\boldsymbol{\theta}} - \boldsymbol{\theta}\right) \cdot \left( \hat{\boldsymbol{\theta}} - \boldsymbol{\theta} \right)^{\dagger} \right\}$ that satisfies\footnote{$\succeq \mathbf{0}$ means that the matrix on the LHS of Eq.~(\ref{eq:CRLB_psd}) is positive semi-definite: $\forall x \in \mathbb{R}^m$, $x^\dagger \cdot (K_{\hat{\boldsymbol{\theta}}} - \mathcal{I}_{\boldsymbol{\theta}}^{-1})\cdot x \geq 0$.}
\begin{equation} \label{eq:CRLB_psd}
K_{\hat{\boldsymbol{\theta}}} - \mathcal{I}_{\boldsymbol{\theta}}^{-1} \succeq \mathbf{0}.
\end{equation}

$\mathcal{I}_{\boldsymbol{\theta}} \in \mathcal{M}^{m \times m}$ denotes the Fisher information matrix, whose components are defined by  $\forall i,j \in \{1,\dots,m \},$
\begin{equation}
[\mathcal{I}_{\boldsymbol{\theta}}]_{(i,j)} 
\equiv -\mathbb{E} \left\{ \frac{\partial^{2} \ln{L(\mathbf{I};\boldsymbol{\theta},\mathcal{N})}}{\partial \theta_{i} \partial \theta_{j}} \right\} \,.
\label{eq:fisher_matrix}
\end{equation}

(ii) Furthermore, if there exists a function $\mathbf{h}: \mathbb{R}^n \to \boldsymbol{\Theta}$ such that 
\begin{equation}
\frac{\partial \ln{L(\mathbf{I};\boldsymbol{\theta},\mathcal{N})}}{\partial \theta_{i}} = \left[ \mathcal{I}_{\boldsymbol{\theta}} (\mathbf{h}(\mathbf{I}) - \boldsymbol{\theta}) \right]_{i} \,, \hspace{3mm} \forall i \in \{1,\dots,m\} \,, \label{eq:CRLB_g_cond}
\end{equation}
then the minimal variance unbiased estimator (MVUE) is given by $\hat{\boldsymbol{\theta}} = \mathbf{h} (\mathbf{I})$, and its optimal covariance matrix is $\mathcal{I}_{\boldsymbol{\theta}}^{-1}$.
\label{theo:CRLB}

\end{theorem}
From Theorem~1, any unbiased estimator $\hat{\theta}_i$ of $\theta_i$ satisfies that (from Eq.~(\ref{eq:CRLB_psd}))
\begin{equation}
Var(\hat{\theta}_{i}) = [K_{\hat{\boldsymbol{\theta}}}]_{(i,i)} \geq [\mathcal{I}_{\boldsymbol{\theta}}^{-1}]_{(i,i)} \,, \hspace{1mm} \forall i \in \{1,\dots,m\}, 
\end{equation}
and therefore $[\mathcal{I}_{\boldsymbol{\theta}}^{-1}]_{(i,i)}$ is a lower bound for the mean squared error (MSE) of $\hat{\theta}_i$.\footnote{Theorem~1 does not guarantee the existence of an estimator that achieves the CRLB; however, it is still possible to find the MVUE \citep{kay1993fundamentals}.}

\subsection{CRLB for the joint estimation of source flux and background} \label{sec:CRLB_analysis}

First, we revisit a result in which the CRLB is derived for the joint source flux and background estimation. We have the following result:

\begin{lemma}\cite[Lemma 3.1]{2024PASP..136a4501V}\label{lm_fisher_joint}
If the astrometry $x_c \in \mathbb{R}$ is fixed and known, and we want to estimate the pair $(F, B)$ from $\mathbf{I} \sim   L(\mathbf{I} ;F, B,\mathcal{N})  $ in (\ref{eq:likelihood}), then the Fisher information matrix in Eq.~(\ref{eq:CRLB_psd}) is given by
\begin{align}
\mathcal{I}_{\boldsymbol{\theta}} & = \begin{bmatrix}
\mathcal{I}_{1,1}(\mathcal{N}) & \mathcal{I}_{1,2}(\mathcal{N}) \\
\mathcal{I}_{2,1}(\mathcal{N}) & \mathcal{I}_{2,2}(\mathcal{N})
\end{bmatrix} & \equiv \begin{bmatrix}
\sum_{i \in \mathcal{N}}\limits \frac{g_{i}^{2}(x_{c})}{\lambda_{i}(F,B)} & \sum_{i \in \mathcal{N}}\limits \frac{g_{i}(x_{c})}{\lambda_{i}(F,B)} \\
\sum_{i \in \mathcal{N}}\limits  \frac{g_{i}(x_{c})}{\lambda_{i}(F,B)} &  \sum_{i \in \mathcal{N}}\limits  \frac{1}{\lambda_{i}(F,B)}
\end{bmatrix} \,.\label{eq:FB_fisher}
\end{align}  
\end{lemma}
Consequently, the following two-dimensional CRLB for any pair of unbiased estimates $\hat{F}$ and $\hat{B}$ can be respectively defined as
\begin{equation}
Var(\hat{F})  \geq \sigma_{F}^{2}  = [\mathcal{I}_{\boldsymbol{\theta}}^{-1}]_{(1,1)}= \frac{\mathcal{I}_{2,2}(\mathcal{N})}{\mathcal{I}_{1,1}(\mathcal{N}) \cdot \mathcal{I}_{2,2}(\mathcal{N}) - \mathcal{I}_{1,2}^{2}(\mathcal{N})}  \,,\label{eq:FBound}
\end{equation}
\begin{equation}
Var(\hat{B}) \geq \sigma_{B}^{2} = [\mathcal{I}_{\boldsymbol{\theta}}^{-1}]_{(2,2)}= \frac{\mathcal{I}_{1,1}(\mathcal{N})}{\mathcal{I}_{1,1}(\mathcal{N}) \cdot \mathcal{I}_{2,2}(\mathcal{N}) - \mathcal{I}_{1,2}^{2}(\mathcal{N})}  \,. \label{eq:BBound}
\end{equation}
We present the equivalent joint Fisher information for the flux and background as
\begin{equation}
\mathcal{I}_{F}(\mathcal{N})  = \mathcal{I}_{1,1}(\mathcal{N})- \frac{\mathcal{I}_{1,2}^{2}(\mathcal{N})}{\mathcal{I}_{2,2}(\mathcal{N})},
\end{equation}
\begin{equation}
\mathcal{I}_{B}(\mathcal{N})  = \mathcal{I}_{2,2}(\mathcal{N})- \frac{\mathcal{I}_{1,2}^{2}(\mathcal{N})}{\mathcal{I}_{1,1}(\mathcal{N})}.
\end{equation}
The statistical interaction (correlation) between components is captured
by the off-diagonal elements of the Fisher information matrix in Eqs.~(\ref{eq:FBound}) and~(\ref{eq:BBound}). Importantly, in the special case that $\mathcal{I}_{1,2} = 0$, the estimates become decoupled and one could consider that the joint estimation task reduces to two isolated 1D estimation problems. 
\subsection{Model assumptions} \label{modelasu}

In this subsection, we outline the assumptions that will aid in the analysis of the photometric errors induced by a given aperture (PSF fitting radius) choice. These assumptions are based on the fact that the observations have good coverage and resolution when capturing the images.

\subsubsection{Good spatial coverage of the PSF on the detector} \label{sec:coverage}

A basic assumption here is that we have a good coverage of the object of interest, in the sense that for a given position $x_c,$ then this condition implies that the set of all $n$ available pixels  is such that
\begin{equation}
\sum_{i \in \mathcal{N}}\limits g_{i}(x_{c}) \approx \int_{-\infty}^{\infty} \phi(x-x_{c},\sigma) dx = 1 \,.\label{eq:good_coverage}
\end{equation}

\subsubsection{High-resolution regime} \label{sec:HR_approx}

This approximation considers that the imaging device allows for a sufficiently high resolution, such that $\Delta x \ll \sigma$. In such a scenario, the following approximation can be considered:
\begin{equation}
g_{i}(x_{c}) \approx \phi(x_{i} - x_{c},\sigma) \cdot \Delta x \,.\label{eq:gi_approx}
\end{equation}
To account for the actual coverage of the chosen aperture, we recognize that the following function from~\citet{2013PASP..125..580M,2014PASP..126..798M} allows us to do that:
\begin{equation}
P(u) \equiv \erf{\left( \frac{u}{\sqrt{2}\sigma} \right)} =\frac{2}{\sqrt{\pi}} \int_{0}^{\frac{u}{\sqrt{2}\sigma}} e^{-v^{2}} dv\,.\label{eq:P_u}
\end{equation}
Here $u$ is the aperture radius measured in the same units as $\sigma$ ({e.g.,} in arcseconds). The quantity $P(u)$ represents the fraction of the signal that is captured within the chosen aperture, which is exactly the interpretation of the more general term $\sum_{i\in\mathcal{J}_u}g_{i}(x_{c})$. 

\subsubsection{Approximations for $g_i^p(x_c)$}

We note that the expressions associated with Fisher information in Eq. (\ref{eq:FB_fisher}) involve discrete sums; however, we can calculate approximations to obtain continuous representations as a function of the aperture radius $u$. The following proposition provides closed-form expressions for the sum of $g_i(x_c)$ across the pixels.

\begin{proposition} \label{prop:approximation}

With the high-resolution assumption, we have the following approximations for the sum of $g_i(x_c)$; in particular, we have that
\begin{equation}
\sum_{i\in\mathcal{J}_u}g_{i}^{p}(x_{c}) \approx \frac{1}{\sqrt{p}} \cdot \left( \frac{1}{\sqrt{2\pi}} \right)^{p-1} \cdot \left( \frac{\Delta x}{\sigma} \right)^{p-1} \cdot P(u\sqrt{p}) \,,
    \label{eq:aperture_sums_p}
\end{equation}
for $p \geq 1$. Analogously (also for $p \geq 1$), it can be found that
\begin{equation}
\begin{split}
    \sum_{i\in\mathcal{J}_u} \frac{1}{g_{i}^{p}(x_{c})}  & \approx  \frac{ \left( \sqrt{2\pi} \right)^{p+1} }{\sqrt{p}}  \left( \frac{\sigma}{\Delta x} \right)^{p+1}  \erfi{\left( \frac{u}{\sqrt{2}\sigma} \sqrt{p} \right)} \,,
\end{split}     
\label{eq:aperture_sum_-p}
\end{equation}
where $\erfi{(x)}$ is defined through the $\erf$ function as
\begin{equation}
\erfi{(x)} = -\iu\erf{(\iu x)} \,, 
\end{equation}
and $\iu = \sqrt{-1}$ corresponds to the imaginary unit.

\end{proposition}

These results are demonstrated in Appendix~\ref{app:HR_sums_powers}. The approximations proposed here are useful as they allow us to derive analytical closed-form expressions for the Fisher information terms and to perform an in-depth analysis, which is presented in the following section. 

\subsection{Jitter and position estimation in the PSF model}

A potential source of photometric performance degradation, in particular in space-based photometry, is satellite jitter, which are small pointing errors occurring during an exposure. These fluctuations modify the effective distribution of photons on the detector, altering the shape and localization of the observed PSF. In this work, we assume that the source centroid $x_c$ is known with high precision throughout the exposure. However, if the effect of jitter is to be included explicitly, the appropriate treatment is to consider a joint estimation problem where both $F$ and $x_c$ are unknown parameters. In such case, the Fisher information matrix includes cross-terms that quantify the coupling between photometry and astrometry, and marginalizing over $x_c$ reduces the information available for estimating the flux. This formalism has been studied in detail by \citet{2014PASP..126..798M}, who conducted a systematic Fisher information analysis to characterize how uncertainty in the source position $x_c$ degrades the precision of photometric estimates, thereby quantifying the impact of any perturbation in source localization, such as those induced by satellite jitter.

\section{Aperture (PSF fitting radius) analysis} \label{sec:aperture}

Photometry plays a crucial role in accurately estimating properties of celestial objects, where the selection of pixels as a function of aperture radius is essential. Classical PSF fitting studies in this field have primarily focused on maximizing the S/N to enhance measurement precision. In this section we propose a new paradigm, focusing instead on the Fisher matrix and in how it can be interpreted as an object containing the necessary information (within a certain quantifiable variance) for the task of aperture selection. First, we  revisit the formal definition of S/N, and based on this, we  compare the aperture selection following the classic S/N-based scheme with our new CRLB-based approach.

\subsection{S/N as a function of the aperture radius}

The S/N is a common metric used to assess the quality of various measurements performed on a celestial object. We revisit the analytical expression obtained from \citet{2013PASP..125..580M} for the S/N and then derive properties that will be useful in the following sections. For the case of the Gaussian function PSF, it makes sense to perform an integration of the PSF that is symmetrical with respect to the center of the source, centered at $x_c$. On the other hand, the total noise, $N$, has contributions from the readout noise of the detector, the noise from the sky, and the noise from the source itself, all of which are assumed to follow Poisson statistics. While we are concerned here with crowded fields, which require a small aperture, we assume, within that selected aperture, that the influence of nearby objects is negligible. Integrating what was previously mentioned, the S/N can be written as a function of the aperture radius (see \citet{1992ASPC...23...68G} and \citet[Eq. (28)]{2013PASP..125..580M}. More specifically, we have the following expression:
\begin{equation}
S/N(u)=\frac{P\left (\sqrt{2}\sigma u \right )F}{\sqrt{P\left (\sqrt{2}\sigma u \right )F+4\sigma^2 u(B-D)}}.
\label{SNRS}
\end{equation}
It is interesting to note that this function has a maximum with respect to $u$. This is because as the aperture radius increases, the noise contribution also increases since the flux eventually saturates and there are more pixels contributing with background and noise. The following proposition summarizes the concavity property of the S/N.

\begin{proposition} \label{propape}
The function $S/N(u)$ has a maximum value $u^{\ast}_{S/N}$, moreover, this optimal aperture radius is given as the solution to the following equation:

\begin{equation}
\begin{split}
 &F \frac{\partial P\left (\sqrt{2}\sigma u^{\ast}_{S/N} \right )}{\partial u}P\left (\sqrt{2}\sigma u^{\ast}_{S/N}\right )= \\
 &-\left [ 2\sqrt{2}\sigma u^{\ast}_{S/N}\frac{\partial P\left (\sqrt{2}\sigma u^{\ast}_{S/N}\right )}{\partial u} - P\left (\sqrt{2}\sigma u^{\ast}_{S/N}\right ) \right ]  2\sqrt{2}\sigma\left (B-D\right ).   
\end{split}
\end{equation}

\end{proposition}
The proof is presented in Appendix \ref{appendSNR}. Proposition 2 tells us that there is a criterion for selecting the aperture radius for photometry based on maximizing the S/N \citep{1989PASP..101..616H,1991PASP..103..122N}.

\subsection{Joint Fisher information as a function of the aperture radius}

We formalize the concept of photometric uncertainty for a given aperture, based on Fisher information, recognizing its role as an optimal performance indicator for unbiased estimators concerning their variance in the context of the CRLB.

The Fisher information plays a crucial role as it acts as a fundamental limit for the joint estimation problem as a function of the number of pixels. In particular, we see that there is a direct connection between Fisher information and the aperture radius, which motivates a systematic analysis regarding the impact of the selected aperture on the associated photometric precision. We restrict our analysis to aperture-like sets of indices, {i.e.,} sets of the form $\mathcal{J}_u~=~\{n_{x_{c}} - n_{u}, \dots, n_{x_{c}}, \dots, n_{x_{c}} + n_{u}\}$, where $n_{u}$ is the number of full pixels within the aperture radius. 

First of all, thanks to the additive structure in Eq.~(\ref{eq:FB_fisher}), we see that the Fisher information in the joint photometric case is given, for an arbitrary set of indices $\mathcal{J}_u$,  by
\begin{align}
    \mathcal{I}_{\boldsymbol{\alpha}^{\star}}(\mathcal{J}_u)  & = \begin{bmatrix}
\mathcal{I}_{1,1}(\mathcal{J}_u) & \mathcal{I}_{1,2}(\mathcal{J}_u) \\
\mathcal{I}_{2,1}(\mathcal{J}_u) & \mathcal{I}_{2,2}(\mathcal{J}_u)
\end{bmatrix} & \equiv   \begin{bmatrix}
        \sum_{i\in\mathcal{J}_u} \limits \frac{g_{i}^{2}(x_{c})}{\lambda_{i}(F,B)} & \sum_{i\in\mathcal{J}_u}\limits \frac{g_{i}(x_{c})}{\lambda_{i}(F,B)} \\
        \sum_{i\in\mathcal{J}_u} \limits \frac{g_{i}(x_{c})}{\lambda_{i}(F,B)} & \sum_{i\in\mathcal{J}_u} \limits \frac{1}{\lambda_{i}(F,B)}
    \end{bmatrix}. 
    \label{eq:Fisher_matrix}
\end{align}

Given a total of $|\mathcal{N}|=n$ pixels, the flux is dominant around the center of the aperture mask, whereas in pixels outside the aperture radius (or far from the center), the background becomes dominant. By focusing on the flux-dominant pixels at the center, we can estimate the accuracy of photometric measurements for a given PSF fitting radius. Then we have the following expressions for the Fisher information:
\begin{equation}
\begin{split}
\mathcal{I}_{F}(\mathcal{J}_u) & = \mathcal{I}_{1,1}(\mathcal{J}_u)- \frac{\mathcal{I}_{1,2}^{2}(\mathcal{J}_u)}{\mathcal{I}_{2,2}(\mathcal{J}_u)}\\
& =  \sum_{i\in\mathcal{J}_u} \frac{g_{i}^{2}(x_{c})}{\lambda_{i}(F,B)} -\frac{\left ( \sum_{i\in\mathcal{J}_u} \frac{g_{i}(x_{c})}{\lambda_{i}(F,B)}  \right )^2}{ \sum_{i\in\mathcal{J}_u} \frac{1}{\lambda_{i}(F,B)}} \,,\label{eq:I_F}    
\end{split}
\end{equation}
\begin{equation}
\begin{split}
\mathcal{I}_{B}(\mathcal{J}_u) & = \mathcal{I}_{2,2}(\mathcal{J}_u) - \frac{\mathcal{I}_{1,2}^{2}(\mathcal{J}_u)}{\mathcal{I}_{1,1}(\mathcal{J}_u)}\\
& =  \sum_{i\in\mathcal{J}_u} \frac{1}{\lambda_{i}(F,B)}  -\frac{\left ( \sum_{i\in\mathcal{J}_u} \frac{g_{i}(x_{c})}{\lambda_{i}(F,B)}  \right )^2}{\sum_{i\in\mathcal{J}_u} \frac{g_{i}^{2}(x_{c})}{\lambda_{i}(F,B)}}\,.
\label{eq:I_B}
\end{split}
\end{equation}
We propose a novel information criterion, noting that the Fisher information is a direct quantitative indicator of the maximum achievable precision for a given set of pixels (or aperture). It is well known that having all pixels available for the estimation process improves performance, achieving the best photometric precision  \citep{2024PASP..136a4501V} in the CRLB sense,  this is the case of isolated targets. In crowded fields, however, only a subset of pixels is chosen to avoid contamination from nearby sources. As a result, the challenge lies in finding an aperture that captures the maximum amount of useful flux while avoiding interference from nearby sources, ensuring accurate photometric and astrometric estimations within these practical limitations. This motivates the need to explore how to select a subset of pixels such that, with an arbitrarily small error, that particular pixel selection ensures that the Fisher information is close to that obtained when using all pixels. This approach optimizes resource use while maintaining a high level of precision in the photometric estimations. 

We propose the following methodology. Formally, given a discrepancy level $\delta >0$, we seek to find the minimum radius $u^{\ast}$ with its corresponding subset $\mathcal{J}_{u^{\ast}} \subseteq \mathcal{N}$ such that
\begin{equation}
\frac{ \mathcal{I}_{F}(\mathcal{N})- \mathcal{I}_{F}(\mathcal{J}_{u^{\ast}})}{\mathcal{I}_{F}(\mathcal{N})} \leq \delta \label{optiV1}
\end{equation}
or, equivalently,
\begin{equation}
\frac{ \inf_{u \in \left \{1,..., \frac{n-1}{2} \right \}}\limits  \mathcal{I}_{F}(\mathcal{J}_u)}{\mathcal{I}_{F}(\mathcal{N})} \geq 1-\delta \label{optiV2}.
\end{equation} \label{sec:useful_approx}
While Eq.~(\ref{optiV1}) can be implemented numerically, it is a discrete problem since it depends on the pixels and not on the aperture radius. Therefore, it is convenient to make an approximation to reframe the optimization problem and transform it into a continuous optimization problem. To achieve this, we recall that this analysis is a function of the set $\mathcal{J}_u$, and we  consider the assumptions from Sect.~\ref{modelasu}, in particular, the high-resolution regime and the approximations for $g_i(x_c)$. Additionally, we note that given $N$ pixels, the pixels closest to the central pixel contain a larger portion of the flux ($F \gg B$). Imposing these conditions for the case of photometry and background estimation is equivalent to making a zero-order approximation of the joint Fisher information. Integrating all of the above, we have the following result.

\begin{theorem} \label{The:theo}
Under the high-resolution scenario and the approximations for $g_i(x_c)$ in Sect.~\ref{modelasu}, the Fisher information for the flux and background can be written as
\begin{equation}
\mathcal{I}_{F}(\mathcal{J}_u)=\frac{1}{F} \left ( P\left ( u\right )  -\frac{2u^2}{\pi \sigma^2 \erfi\left( \frac{u}{\sqrt{2}\sigma}\right )}  \right ),
\end{equation}

\begin{equation}
\mathcal{I}_{B}(\mathcal{J}_u) = \frac{1}{F} \left (  \left ( \frac{\sqrt{2\pi}\sigma}{\Delta x} \right )^2\erfi\left( \frac{u}{\sqrt{2}\sigma}\right )  -\frac{4u^2}{P(u)(\Delta x)^2 }  \right ).
\end{equation}
Importantly, we have that $\lim_{u\rightarrow \infty}\limits \mathcal{I}_{F}(\mathcal{J}_u)$ exists, which implies that $\lim_{ u \rightarrow \infty} \limits \frac{\partial \mathcal{I}_{F}(u)}{\partial u}=0$.
\end{theorem}
The proof is presented in Appendix~\ref{Appen:Theo}. Then, the optimization formulation in Eq.~(\ref{optiV2}) can be rewritten as
\begin{equation}
u^{\ast}_{\mathcal{I}_{F}} =  \arg \min_{u \in \mathbb{R}^+}\limits  \left ( P\left ( u\right )  -\frac{2u^2}{\pi \sigma^2 \erfi\left( \frac{u}{\sqrt{2}\sigma}\right )}  \right )
\label{optiI}
\end{equation}
such that $\frac{\left ( P\left ( u^{\ast}_{\mathcal{I}_{F}} \right )  -\frac{2\left (u^{\ast}_{\mathcal{I}_{F}} \right )^2}{\pi \sigma^2 \erfi\left( \frac{u^{\ast}_{\mathcal{I}_{F}} }{\sqrt{2}\sigma}\right )}  \right )}{ \left ( P\left ( u_{max}\right )  -\frac{2u_{max}^2}{\pi \sigma^2 \erfi\left( \frac{u_{max}}{\sqrt{2}\sigma}\right )}  \right )} \geq 1-\delta$.

\begin{remark}
Theorem~2 tells us that both the joint Fisher information in flux and background have a monotonic behavior as a function of the aperture radius. This is important because it reinforces the results presented in \citet{2024PASP..136a4501V}; specifically, and as expected, to achieve higher precision in estimation, it is necessary to use all available pixels. Any other method based on processing a subset of the available observations will inevitably lead to a loss of information. This is because, according to the data processing inequality \citep{cover1999elements}, the information about the parameter decreases with each processing step.
\end{remark}

\begin{remark}
 Equation~(\ref{optiV1}) tells us that we can be arbitrarily close to the value that maximizes Fisher information by choosing an appropriate delta. However, the contribution per pixel gradually decreases, leading to a saturation zone. This feature allows for an aperture design by identifying the point where such saturation occurs. The proposed criterion provides flexibility as it allows for the adjustment of tolerance values based on the cost (or convenience) of incorporating more pixels. By doing so, we can determine the gain (or loss) in precision when additional pixels are included (or excluded) in the aperture depending on crowding. In practice, this scheme allows for the free selection of a tolerance level by using a larger number of pixels, which can be costly or inadequate (e.g., in a crowded field). Therefore,  the flexibility provided by this method enables the astronomer to conduct a cost-benefit analysis to achieve the highest attainable precision given the observational constraints while using the fewest pixels necessary within the acceptable error budget of the respective science case.
\end{remark}

\section{Numerical results} \label{numerical}

In this section, we present numerical results, focusing on two key experiments. The first experiment examines the variation in Fisher information as a function of the aperture radius, and provides insights into how the choice of this radius impacts  the information available for the joint estimation. The second experiment analyzes the performance of classical astrometric estimators as a function of the aperture radius, comparing their accuracy and precision under different aperture configurations in comparison with the expected CRLB for that aperture choice.

\subsection{Discrepancy analysis as a function of the aperture radius}

Here we present a numerical analysis that lends support to the methodology proposed in Eq. (\ref{optiV1}) and the theoretical expressions derived from Theorem~2. We use Eq. (\ref{optiV1}) to compute, for a selection of the PSF fitting radius, the Fisher information comparing its predictions to the aperture that maximizes the S/N in Eq. (\ref{SNRS}). To this end, we developed various experiments with different flux and background levels. The experimental setup (source, device, and sky conditions) considered here and in the forthcoming sections, which we denote as the baseline case, is summarized in Table~\ref{tab:baseline}. The parameter $dither$ denotes an offset between the center of the $n_{x_{c}}^{th}$ pixel and the actual position of the point source $x_{c}$ \citep{2013PASP..125..580M,2014PASP..126..798M}, measured in pixels.

\begin{table}[htp]
\centering
\caption{Parameters adopted for the baseline case.} \label{tab:baseline}
\begin{tabular}{ccc}
\hline \hline 
Parameter  & Value & Units \\
\hline
$D$     &  0 & [photo-e$^{-}$] \\
RON & 5 & [photo-e$^{-}$] \\
$\Delta x$ & 0.2 & [arcsec] \\
$N$ & 51 & [pixels] \\
$f_s$ & 3005 & [photo-e$^{-}$ arcsec$^{-1}$] \\
$n_{x_c}$ & 26 & - \\
dither & 0 & [pixels] \\
FWHM  & 1.0 & [arcsec] \\
\hline
\end{tabular}
\end{table}
\begin{figure*}
\sidecaption
\begin{minipage}{12cm} 
   \includegraphics[width=\linewidth]{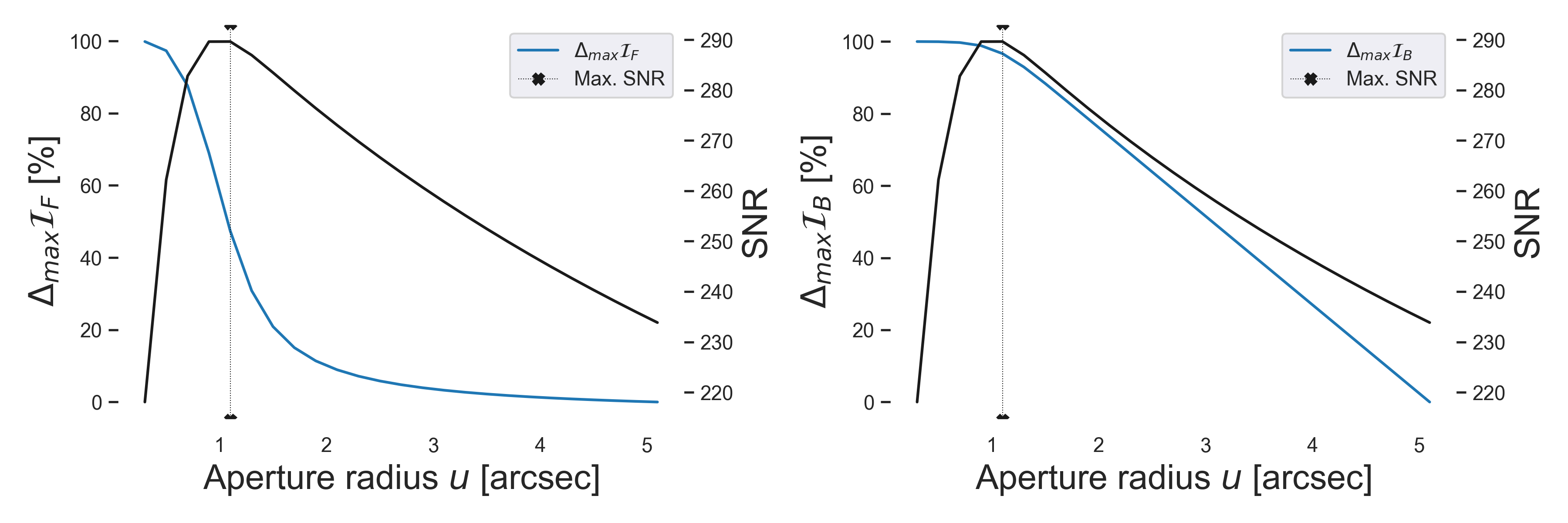}
   \centerline{(a) $F$: 100000 [photo-e$^{-}$]  -  $B$: 1625.0 [photo-e$^{-}$] FWHM: 1.0 [arcsec]}
   \includegraphics[width=\linewidth]{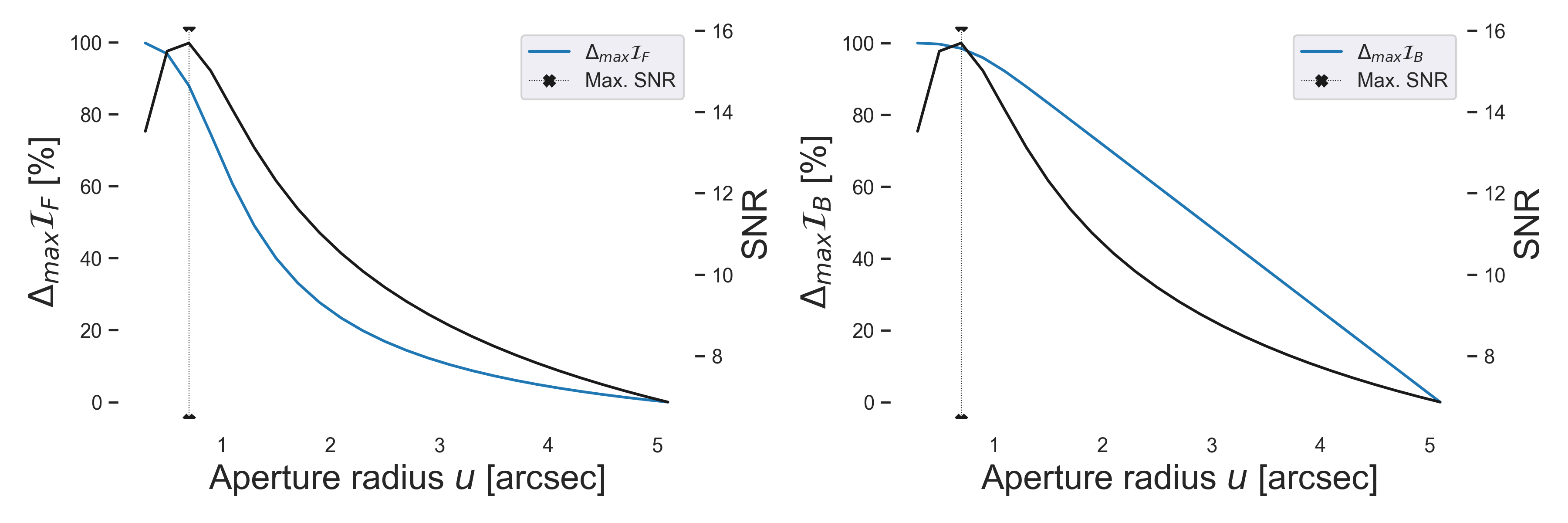}
   \centerline{(b)   $F$: 2000 [photo-e$^{-}$]  -   $B$: 1625.0 [photo-e$^{-}$] FWHM: 1.0 [arcsec]}
   \includegraphics[width=\linewidth]{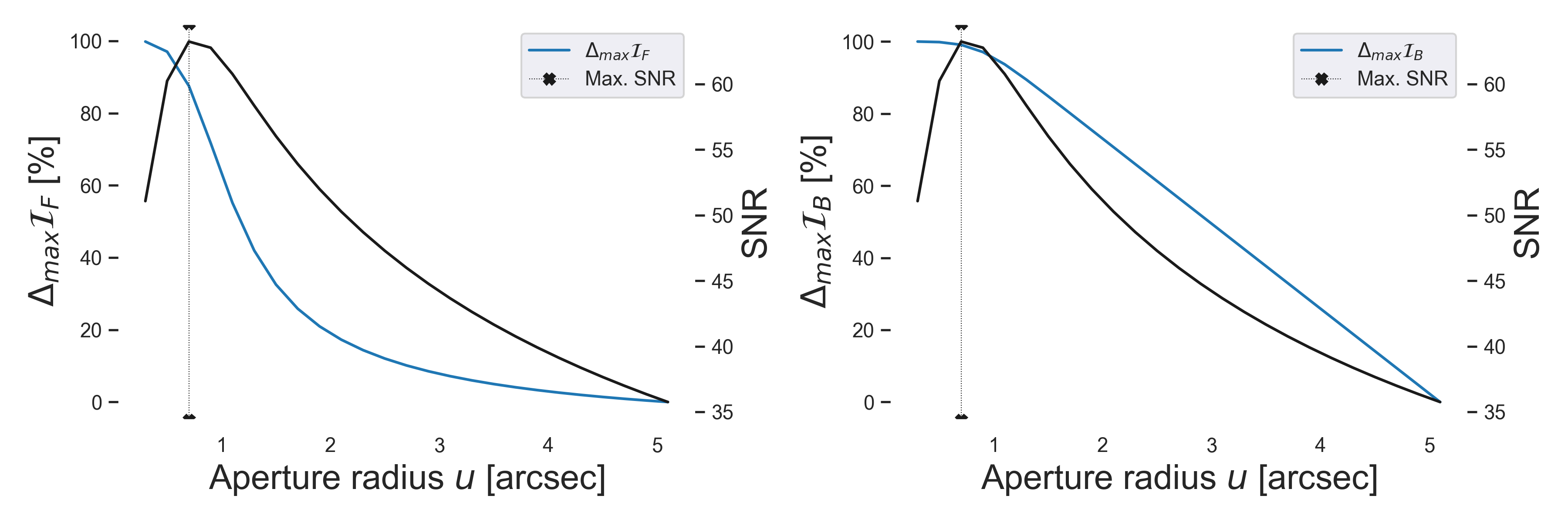}
   \centerline{(c) $F$: 8000 [photo-e$^{-}$]  -  $B$: 825.0 [photo-e$^{-}$] FWHM: 1.0 [arcsec]}
   \includegraphics[width=\linewidth]{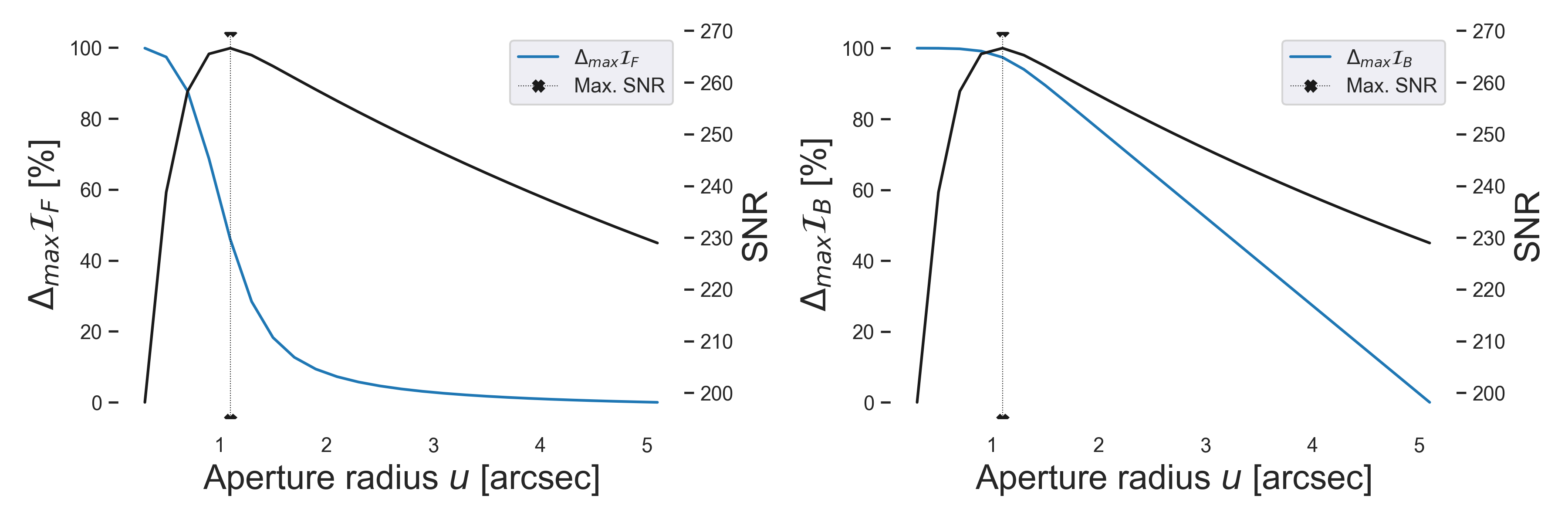}
   \centerline{(d)  $F$: 80000.0 [photo-e$^{-}$]   -   $B$: 825.0 [photo-e$^{-}$] FWHM: 1.0 [arcsec]}
\end{minipage}
\caption{Discrepancy analysis of the joint Fisher information as a function of the PSF aperture fitting radius in photometry (blue line, left ordinate on each plot, in \%). The discrepancy in the source flux (left column) is defined as $\Delta_{max}\mathcal{I}_F=\frac{ \mathcal{I}_{F}(\mathcal{N})- \mathcal{I}_{F}(\mathcal{J}_{u})}{\mathcal{I}_{F}(\mathcal{N})} $. The discrepancy in the background (right column) is defined as $\Delta_{max}\mathcal{I}_B=\frac{ \mathcal{I}_{B}(\mathcal{N})- \mathcal{I}_{B}(\mathcal{J}_{u})}{\mathcal{I}_{B}(\mathcal{N})}$. The black line (right ordinate on each plot) indicates the corresponding S/N value. The vertical line shows the aperture at which the S/N is maximized. Results are reported for different representative values of $F, B$, and a FWHM of 1.0~arcsec.}
 \label{fig5:a}
\end{figure*}
For example, for a flux of 10,002 [photo-e$^{-}$], we obtain an S/N of 78.\footnote{This flux was chosen for illustration purposes so that, with the other parameters in this example, the resulting S/N is exactly 78.} The first experiment involves analyzing the evolution of the discrepancy in the joint CRLB for the photometric case as the aperture radius increases. Specifically, by discrepancy, we mean that for a given PSF fitting (aperture) radius $u$, we compute $\frac{ \mathcal{I}_{F}(\mathcal{N})- \mathcal{I}_{F}(\mathcal{J}_{u})}{\mathcal{I}_{F}(\mathcal{N})} $ for the source flux, and $\frac{ \mathcal{I}_{B}(\mathcal{N})- \mathcal{I}_{B}(\mathcal{J}_{u})}{\mathcal{I}_{B}(\mathcal{N})} $ for the background. We assess how far the (variance) bound deviates from the value obtained when considering all pixels. Additionally, the S/N is plotted as a function of the aperture radius, indicating the point where its value is maximized. Figure~\ref{fig5:a} shows the results of this experiment for different configurations of $F$ and $B$.

In terms of the source flux, as expected, we observe a monotonic decrease in the CRLB variance (i.e., increasing precision); as the aperture radius increases, more information from the source is involved, leading to a reduction in the variance associated with the flux estimation. The discrepancy also decreases when the difference between the flux $F$ and the background $B$ increases, which is expected. A higher flux value relative to the background means a cleaner signal, thereby improving photometric precision. On the other hand, a larger FWHM implies a flux signal spread wider across the instrument, which results in a slower rate at which the best precision is reached when all pixels are available. Moreover, we note that the aperture radius that maximizes the S/N corresponds to a discrepancy percentage (in the terms defined previously) between 40\% and 85\%.

\begin{table*}[ht]
\centering
\caption{Comparison of discrepancies as a function of the PSF fitting radius.} 
\label{tab:aperture_comparison}
\begin{tabular}{ccccccc}
\hline 
Figure & Radius at Max S/N & Radius at 1.2$\times$FWHM & Radius at 10\% Discrepancy  & $F$& $B$ & $F/B$ \\
 & / Discrepancy & / Discrepancy & / Discrepancy  & e$^-$  & e$^-$ &\\
\hline
--- & 0.8" / 48\% &  0.6" / 44\% & 1.5" / 10\%  & 20000 & 2025 & 9.87\\
--- & 0.4" / 92\% &  0.6" / 78\% & 2.1" / 10\% & 2000 & 1025 & 1.95\\
--- & 0.4" / 91\% &  0.6" / 78\% & 2.7" / 10\% & 100 & 1000 & 0.10\\
--- & 0.6" / 39\% &  0.6" / 39\% & 0.8" / 10\% & 100000 & 1025 & 97.56\\
\hline
(a) & 1.08" / 45\% &  1.2" / 38\% & 2.09" / 10\% & 100000 & 1625 & 61.54 \\
(b) & 0.69" / 85\% &  1.2" / 56\% & 3.30" / 10\% & 2000 & 1625 &1.23 \\
(c) & 0.70" / 85\% &  1.2" / 46\% & 2.80" / 10\% & 8000 & 825 &9.69 \\
(d)& 1.08" / 44\%  &  1.2" / 37\% & 1.84" / 10\% & 80000 & 825 &96.96\\
\hline
--- & 1.58" / 56\% &  1.8" / 28\% & 2.8" / 10\% & 100000 & 1625 & 61.54\\
--- & 0.90" / 85\% &  1.8" / 47\% & 3.7" / 10\% & 2000 & 1625 & 1.23 \\
--- & 1.10" / 85\% &  1.8" / 44\% & 3.8" / 10\% & 10000 & 1625 & 6.15 \\
--- & 1.58" / 55\% &  1.8" / 36\% & 2.8" / 10\% &80000 & 825 & 96.97\\
\hline
\end{tabular}
\tablefoot{ A fitting radius close to the maximum S/N is advocated by \citet{1987PASP...99..191S}[Section III.D] in DAOPHOT, as well as \citet{1998MNRAS.296..339N} (second column). A slightly larger radius is suggested by \citet{mendez2010}[Fig. 4 in Section 3.2] (for astrometry, third column). As a reference, the last column is for a discrepancy tolerance of 10\%. The first four rows are for a FWHM of 0.5~arcsec, the following four rows are for 1.0~arcsec (panels~(a) to (d) in Fig.~\ref{fig5:a}), while the last four are for 1.5~arcsec.}
\end{table*}
Table \ref{tab:aperture_comparison} presents the aperture for several popular choices of the PSF fitting radius for a FWHM of 0.5, 1.0 (baseline case, see Table~\ref{tab:baseline}), and 1.5~arcsec, and its corresponding discrepancy values. These findings suggest that there is room for improvement in the joint estimation of background and flux if a more appropriate radius is selected, for example one that limits the discrepancy to a pre-specified discrepancy (e.g., 10$\%$).

In the case of the background, we observe that there is no sharp decline when the aperture radius increases. Obviously, the flux is dominant around the center of the pixel, whereas in pixels outside the aperture radius, the background becomes linearly dominant. Pixels outside the aperture radius are, therefore, more informative for the background than for the flux. This observation motivates the selection of pixels for determining the aperture radius: the background pixels should complement those selected for the flux. By focusing on the flux-dominant pixels in the center and the background-dominant pixels outside the aperture, we can optimize the aperture radius to enhance the accuracy of photometric measurements. This complementary selection ensures a more precise distinction between the flux and background, thereby improving the reliability of the estimates. This is explored in the next subsection.

\subsubsection{Discrepancy analysis as a function of the halo}

\begin{figure*}
\centering
\includegraphics[width=17cm]{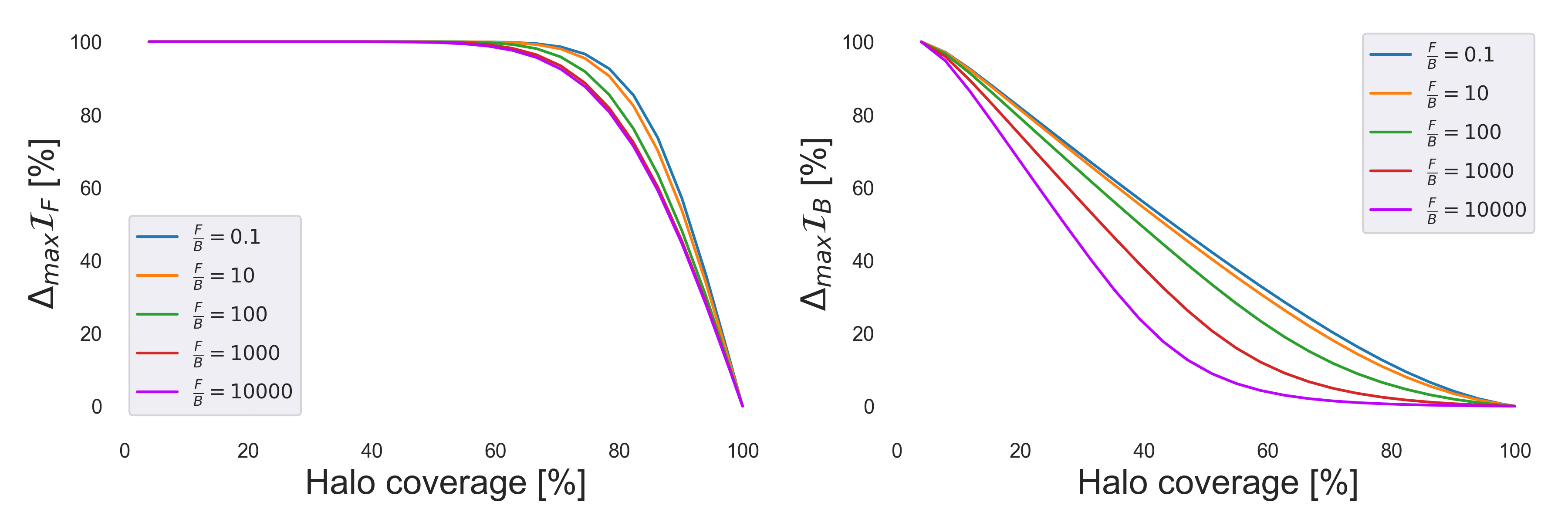} 
\caption{Discrepancy analysis of the joint Fisher information (in \%) as a function of the halo coverage in photometry. The left panel is for the source flux; the right panel is for the background. Results are reported for different representative values of $F/ B$.}
\label{fig6:a} 
\end{figure*}

To further the analysis presented above, we consider the complement set of $\mathcal{J}_u$, which includes the pixels located outside the aperture radius. This set corresponds to a halo, as it consists of pixels outside the central region and symmetrically distributed around it. Figure~\ref{fig6:a} presents an inverse analysis, showing the discrepancy as the halo encompasses more pixels until it includes all of them. Interestingly, achieving a more pronounced decay in the discrepancy requires a higher flux-to-background ratio, $F/B$. This relationship highlights the critical role of the $F/B$ ratio in optimizing the precision and effectiveness of photometric analyses, particularly when considering contributions from regions beyond the immediate aperture. The Fisher information typically increases for the background estimation with a higher S/N because a stronger signal enhances the ability to disentangle the source flux from the background. Intuitively, one might expect the opposite—that a lower S/N would emphasize the background—because noise dominates in such conditions. However, the relationship is nuanced. When the S/N is high, the observed flux is more distinct from the background, making it easier to accurately model and estimate the contribution from  background. This clarity increases the Fisher information for the background because the data provides more reliable clues about the background's statistical behavior. In contrast, when the S/N is low, the noise dominates, and the background becomes harder to distinguish from the signal. This uncertainty reduces the Fisher information, as the data carries less discernible information about the background parameters.

\subsubsection{Impact of contaminating stars on the aperture optimization}
Recent advances in stellar photometry, particularly from \emph{Gaia}, have made it possible to precisely know the positions of neighboring stars in the field and to estimate their fluxes with high accuracy. This enables the explicit incorporation of contaminating stars into the photometric model for aperture optimization.
In the standard framework, the expected counts in pixel $i$ is modeled as in Eq. (\ref{eq_pre_2b}). This assumes that the background is spatially uniform. When contaminating stars are present, the background becomes spatially variable and can be modeled as
\begin{equation}
\tilde{B}_i = B + \sum_{k=1}^{K} F_k\, g_{ik}(x_{ck}),
\end{equation}
where $B$ is the constant background level, $F_k$ is the flux of contaminating star $k$, and $g_{ik}(x_{ck})$ is the PSF of star $k$ integrated over pixel $i$, shifted according to its position relative to the target. The total expected count is therefore
\begin{equation}
\lambda_i(F,\tilde{B}_i) = F\, g_i(x_{ck}) + \tilde{B}_i .
\end{equation}
In this scenario, $B_i$ is assumed to be known from \emph{Gaia}-based positions and pre-measured or modeled fluxes, and only $F$ is to be estimated. The Fisher information for $F$ simplifies to
\begin{equation}
I_{1F}(\mathcal{J}_u) = \sum_{i \in \mathcal{J}_u} \frac{g_i^2(x_c)}{F g_i(x_c) + \tilde{B}_i}.
\end{equation}
Unlike the constant-background case, the background $B_i$ is a fixed known function of pixel position.

For aperture optimization, the discrepancy at aperture radius $u$ is defined as
\begin{equation}
\mathrm{disc}(u) = 1 - \frac{I_{1F}(\mathcal{J}_u)}{I_{1F}(\mathcal{N})},
\label{normEQ}
\end{equation}
where $I_{1F}(\mathcal{J}_u)$ is computed using only pixels within radius $u$ from the target’s center, and $I_{1F}(\mathcal{N})$ is the Fisher information using the full set of pixels in the analysis window. This discrepancy quantifies the fractional loss of Fisher information when restricting the aperture to a given radius.

\begin{figure*}[t]
    \centering
    \includegraphics[width=17cm]{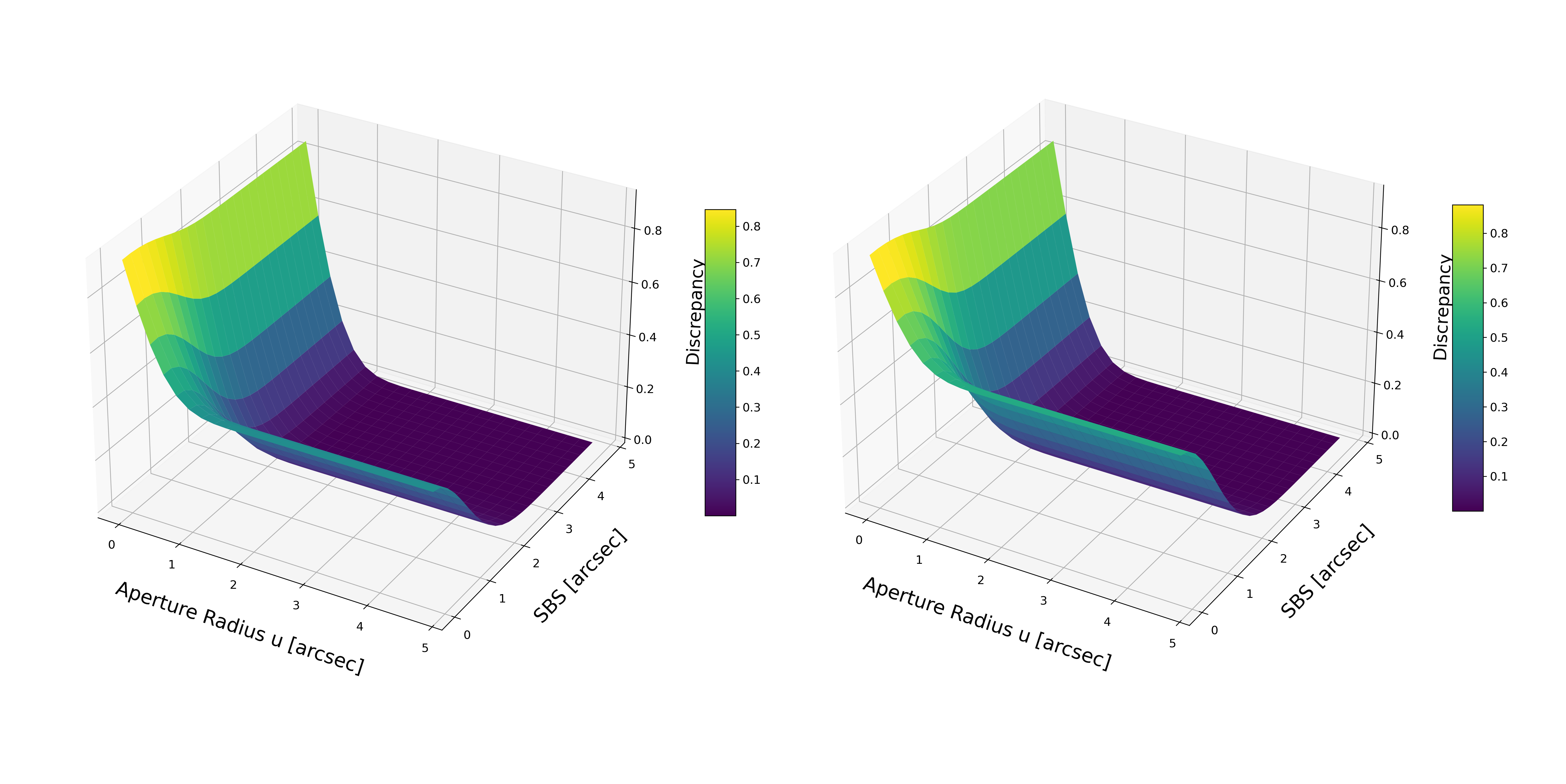}
    \caption{
    Normalized discrepancy (vertical axis, Eq. (\ref{normEQ})) as a function of aperture radius and separation between the contaminating star (SBS) and the target (horizontal axis), for two contamination scenarios. 
    The left panel corresponds to a faint contaminant ($F=2000$ [photo-e$^{-}$], $F_K=1000$ [photo-e$^{-}$], $K=1$), while the right panel shows the effect of a brighter one ($F=2000$ [photo-e$^{-}$], $F_K=2500$ [photo-e$^{-}$], $K=1$). 
    Colors indicate the fractional loss of Fisher information relative to the full-pixel case. }
    \label{fig:fisher_multiple}
\end{figure*}

To quantify the impact of contamination from nearby stars, the background $B_i$ was modeled including one or more contaminating sources at varying separations from the target. Each contaminant was assumed to share the same Gaussian PSF as the target, with a fixed FWHM, and its flux contribution was scaled according to its total flux $F_k$. The separation between the target and the contaminant was systematically varied from zero to several arcseconds. For each separation, the discrepancy was computed as a function of aperture radius. 

The results are shown in Figure~\ref{fig:fisher_multiple}, where the horizontal axis represents both the aperture radius and the separation between the contaminating star (SBS) and the target, while the vertical axis displays the color scale of the normalized discrepancy.
The left panel of Figure~\ref{fig:fisher_multiple} corresponds to the case of a faint contaminant ($F=2000$ [photo-e$^{-}$], $F_K=1000$ [photo-e$^{-}$], $K=1$), while the right panel illustrates the effect of a brighter one ($F=2000$ [photo-e$^{-}$], $F_K=2500$ [photo-e$^{-}$], $K=1$). In both cases, a clear trade-off emerges: at small separations, when the contaminant lies within the core of the PSF, the discrepancy remains high across all aperture sizes. This indicates a significant loss of Fisher information due to the strong overlap between the target and contaminant. As the separation increases, the contamination effect decreases, and the discrepancy curves approach those of the uncontaminated case. For large separations, the contamination is negligible and the optimal aperture radius converges to the same value as in the constant-background scenario.

It is worth noting that, from the perspective of the model, the effect of a single bright contaminant located close to the target can be interpreted as equivalent to the combined effect of multiple fainter stars at similar distances since what ultimately matters is the cumulative flux contribution of contaminants to the pixel $i$ through the term $B_i = B + \sum_{k=1}^{K} F_k\, g_{ik}(x_{ck})$. Consequently, the analysis of the aperture optimization problem in the presence of multiple nearby stars reduces to understanding how these additional flux terms modify the spatial profile of $B_i$ and, therefore, the Fisher information curve as a function of aperture radius.

For the purposes of this work, we restricted our attention to the case of a single well-isolated source, which allowed us to focus on the fundamental behavior of the methodology without the added complexity of crowding effects. A detailed treatment of how nearby sources impact photometric precision and the choice of the optimal aperture will be the subject of future work.

\subsection{Discrepancy analysis for practical high dimensional implicit estimators}

We now analyze the behavior of some classical estimators used in photometry for flux estimation as a function of the aperture radius. Given the equivalence between aperture radius and the number of pixels used, it is important to analyze how much precision is lost when fewer observations are employed. In these numerical experiments, we  examine how much information is available depending on the aperture selection criterion, comparing the use of the S/N-based approach versus the Fisher information-based approach.
First, consider an estimator, or regression rule $\tau_J: \mathbb{N}^{n} \to \boldsymbol{\Theta}$, $\tau_{J}(\cdot)~=~[\tau_{J,1}(\cdot) ,...,\tau_{J,p}(\cdot)]$, which is defined by the solution of
\begin{equation}
\tau_{J}(\mathbf{I}) \equiv \underset{\boldsymbol{\alpha} \in \boldsymbol{\Theta}}{\operatorname{argmin}} J(\boldsymbol{\alpha}, \mathbf{I}), \label{eq:implicit_estimator}
\end{equation}
where $J(\boldsymbol{\alpha}, \mathbf{I})$ is a general cost function. From this point forward, we  consider two classical estimators commonly used in astrometry, along with their associated cost functions: the maximum likelihood estimator and the stochastic weighted least squares estimator. Both estimators have been formally analyzed in detail, and provide theoretical guarantees of optimality in a broad sense across various astrometric and photometric scenarios (see \citet{2015PASP..127.1166L,2018A&A...616A..95E,2024PASP..136a4501V}).

The stochastic weighted least squares estimator is given by the following cost function:
\begin{equation}
J_{SWLS}(\boldsymbol{\alpha}, \mathbf{I}) = \sum_{i =1}^n \frac{1}{I_{i}}(I_{i} - \alpha_{1} \cdot g_{i}(x_{c}) - \alpha_{2})^{2} \,.\label{eq:SWLS_cost}
\end{equation}
This is a particular case of weighted least squares where the selection of the weights can be seen as a noisy version of $\mathbb{E} \left\{ I_i \right\}=\lambda_{i}(F,B)$. On the other hand, the ML estimator has been widely used for parameter estimation due to its asymptotic optimality \citep{kay1993fundamentals}. For this reason, it has been proposed and used by the astronomical community in photometry and astrometry \citep{1993IAUS..156..113C,2013PASP..125..444G,lindegren2008,2017PASP..129e4502G,2018A&A...616A..95E}, and in many other research areas facing similar inverse image problems, such as  fluorescence microscopy \citep{Abraham2009, superresolution2014}.
The likelihood function of $\mathbf{I}$ given the source's relative position $x_{c}$ and its photometry $\boldsymbol{\alpha} \in \boldsymbol{\Theta}$ (see Eq. (\ref{eq:likelihood})), can be expressed as
\begin{equation}
L(\mathbf{I};\alpha,\mathcal{N}) = \prod_{i =1}^n \frac{e^{-\left( \alpha_{1} \cdot g_{i}(x_{c}) + \alpha_{2} \right)} \cdot \left( \alpha_{1} \cdot g_{i}(x_{c}) + \alpha_{2} \right)^{I_{i}}}{I_{i}!} \,.
\end{equation}
Then, the ML estimator is given by
\begin{equation}
\tau_{ML}(\mathbf{I}) = \underset{\boldsymbol{\alpha} \in \boldsymbol{\Theta}}{\operatorname{argmin}} \underbrace{J_{ML}(\boldsymbol{\alpha}, \mathbf{I})}_{= -\ln{L(\mathbf{I};\alpha,\mathcal{N})}}.
\label{eq:ML_tau}
\end{equation}
Using these estimators, the second experiment involves quantifying the change in performance of these estimators as a function of the aperture radius. The aim is to see how the use of fewer pixels or a smaller aperture radius affects the precision of these estimators. Figure~\ref{fig:implicit} (see Appendix \ref{Appen:Fig}) shows the numerical results where we analyze the empirical variance of each estimator and compare it with their variance when all pixels are available. The black vertical line represents the aperture radius associated with the value that maximizes the S/N, while the blue vertical line corresponds to the radius selection based on the Fisher information (i.e., the CRLB), allowing a discrepancy of 20$\%$ according to Eq. (\ref{optiV1}). The green vertical line corresponds to the value of the aperture radius following the approximated methodology in Eq.~(\ref{optiI}). We compared the SWLS and ML estimators as a function of the aperture radius. More precisely, we analyzed the following merit function (expressed as \% of discrepancy with respect to the case of the full pixel\footnote{While this comparison is based on the empirical variance, the result is practically indistinguishable from the theoretical CRLB, as we have demonstrated \citep{2018A&A...616A..95E,2024PASP..136a4501V} that the estimator's performance exhibits a variance close to this bound.}); for visualization, the variances of the source’s flux are expressed in terms of magnitudes \citep{2014PASP..126..798M}:
\begin{equation} 
\sigma_m^2(\mathbf{I}) = \frac{2.5}{2} \cdot \log\left( F+\sqrt{Var(\tau_{J,F}(\mathbf{I}))} \right) - \log\left( F-\sqrt{Var(\tau_{J,F}(\mathbf{I}))}\right),
\end{equation} 
\begin{equation} \label{eq:discr_f}
\Delta_{max} \hat{\sigma}_{m}(u) = \frac{\sigma_m^2(\mathbf{I}_u) -\sigma_m^2(\mathbf{I}) }{\sigma_m^2(\mathbf{I}) } \cdot 100,
\end{equation}

\begin{equation} \label{eq:discr_b}
    \Delta_{max} \hat{\sigma}_{B}(u) = \frac{\sqrt{Var(\tau_{J,B}(\mathbf{I}_u))}-\sqrt{Var(\tau_{J,B}(\mathbf{I}))}}{\sqrt{Var(\tau_{J,B}(\mathbf{I}))}} \cdot 100.
\end{equation}
Here $\tau_{J,F}(\cdot)$ and $\tau_{J,B}(\cdot)$ denote the estimated values of the source flux and background, respectively; $J$ represents the cost function of each estimator,  namely stochastic weighted least squares or maximum likelihood; and $\mathbf{I}_u$ denotes the observations located within the aperture radius $u$. 
As anticipated, we can see that the discrepancy significantly decreases as the aperture radius increases. This result is consistent with the findings from the Fisher information case. Thus, in empirical terms, the performance comparison of the estimators aligns with the fundamental limits. We also note that, for these scenarios, the discrepancy can reach up to 180$\%$ when using the aperture that maximizes the S/N, indicating that the estimation process could be significantly improved if Fisher information is considered. In such a case, the empirical variance would be much closer to the optimal value when more pixels are used.
\section{Conclusions} \label{sec:conclusions}

This research underscores the relevance of using the Fisher information matrix to significantly enhance photometric analysis through an optimal (in the CRLB sense) PSF fitting radius (aperture selection) in crowded fields. By calculating the joint Fisher information matrix for photometric measurements as a function of the aperture radius, we gain valuable insights into how different aperture radii influence the precision of source flux estimates. The joint Fisher information effectively quantifies the amount of information that each pixel or measurement point within an aperture contributes to the final photometric estimate, enabling a more robust data-driven approach to defining the aperture radius. 

Our findings show that while increasing the aperture radius improves the joint Fisher information, and thus enhances the accuracy of flux measurements, there is a practical trade-off. Larger apertures incorporate more pixels, increase computational costs, and eventually run into crowding problems. Consequently, the framework developed here allows for an informed balance between aperture size and the precision of flux estimation, enabling the selection of an optimal aperture that minimizes the loss of information from fewer pixels without unnecessarily inflating computational requirements or incurring in crowding issues. The results suggest that selecting a more appropriate aperture radius (e.g, one that limits the discrepancy to pre-specified percentage) can lead to improvements in the flux estimation process. In  some cases, using S/N-based criteria results in discrepancies as large as 80\% compared to choosing a larger aperture radius. This indicates that a more precise radius selection can significantly reduce estimation errors and improve the reliability of the photometric measurements.

The results presented in this paper are particularly relevant, considering that the joint estimation photometric methodology presented in \citet{2024PASP..136a4501V} assures variances that are very close to the CRLB and that can be attained by practical estimators (as demonstrated by our realistic numerical simulations). Hence, the Cramér-Rao limit is a meaningful benchmark to provide a prescription for the optimal aperture in photometry.

As seen in the numerical results presented in this paper, when the aperture radius is too small, valuable information from the outer pixels of the source’s light distribution is neglected, potentially decreasing the estimator’s precision of flux determination. This methodological advancement shifts photometric practices from traditional, often heuristic, aperture choices to a statistically optimized framework that can adapt based on the source characteristics, background conditions, crowding conditions, and instrumental setup. Finally, this approach ultimately enables more accurate and reliable photometric measurements, providing a powerful tool for precision astronomy in studies ranging from exoplanet detection to deep-sky observations.

\begin{acknowledgements}
SAET, JFS, and MO acknowledge support from the Advanced Center for Electrical and Electronic Engineering (AC3E), funded by the Chilean ANID Basal Project (AFB240002). Additionally, JFS acknowledges support from Fondecyt Project 1250098. RAM acknowledges support from FONDECYT/ANID grant \# 124 0049 and from ANID, Fondo GEMINI, Astrónomo de Soporte GEMINI-ANID grant \# 3223 AS0002.
\end{acknowledgements}
\bibliographystyle{aa} 
\bibliography{references}
\clearpage
\newpage
\begin{appendix}

\section{Proof of Proposition 1}
\label{app:HR_sums_powers}
Equation~\eqref{eq:aperture_sums_p} is a natural consequence of~\eqref{eq:gi_approx} with a Gaussian PSF and considering a set of indices of the form $\mathcal{J}_u~=~\{n_{x_{c}} - n_{u}, \dots, n_{x_{c}}, \dots, n_{x_{c}} + n_{u}\}$:
\begin{equation}
\begin{split}
    & \sum_{i\in\mathcal{J}_u}g_{i}^{p}(x_{c}) \\
    &\approx \sum_{i\in\mathcal{J}_u} \phi^{p}(x_{i} - x_{c},\sigma) \cdot \Delta x^{p} \\
    &= \left( \frac{1}{\sqrt{2\pi}\sigma} \right)^{p} \cdot \Delta x^{p-1} \cdot \sum_{i\in\mathcal{J}_u} \exp{\left( -p \cdot \frac{(x_{i} - x_{c})^{2}}{2\sigma^{2}} \right)} \cdot \Delta x \\
    &\approx \left( \frac{1}{\sqrt{2\pi}\sigma} \right)^{p} \cdot \Delta x^{p-1} \cdot \lim_{\Delta x \to 0} \sum_{i\in\mathcal{J}_u} \exp{\left( -\frac{(x_{i} - x_{c})^{2}}{\frac{2}{p}\sigma^{2}} \right)} \cdot \Delta x \\
    &\approx \left( \frac{1}{\sqrt{2\pi}\sigma} \right)^{p} \cdot \Delta x^{p-1} \int_{x_{c} - u}^{x_{c} + u} \exp{\left( -\frac{(x - x_{c})^{2}}{\frac{2}{p}\sigma^{2}} \right)} dx \\
    &= \left( \frac{\Delta x}{\sqrt{2\pi}\sigma} \right)^{p-1} \cdot \frac{1}{\sqrt{2\pi}\sigma} \cdot 2 \int_{0}^{u} \exp{\left( -\frac{x^{2}}{\frac{2}{p}\sigma^{2}} \right)} dx \\
    &= \left( \frac{\Delta x}{\sqrt{2\pi}\sigma} \right)^{p-1} \cdot \frac{1}{\sqrt{p}} \cdot \frac{2}{\sqrt{\pi}} \int_{0}^{\frac{u}{\sqrt{2}\sigma}\cdot\sqrt{p}} \exp{\left( -x^{2} \right)} dx \\
    &= \frac{1}{\sqrt{p}} \cdot \left( \frac{\Delta x}{\sqrt{2\pi}\sigma} \right)^{p-1} \cdot \erf{\left( \frac{u}{\sqrt{2}\sigma} \sqrt{p} \right)} \\
    &= \frac{1}{\sqrt{p}} \cdot \left( \frac{1}{\sqrt{2\pi}} \right)^{p-1} \cdot \left( \frac{\Delta x}{\sigma} \right)^{p-1} \cdot P(u\sqrt{p}) \,.
\end{split}
\end{equation}
Following similar steps, the approximation of sums of negative powers of $g_{i}(x_{c})$ can be found:
\begin{equation}
\begin{split}
    & \sum_{i\in\mathcal{J}_u} \frac{1}{g_{i}^{p}(x_{c})} \\
    &= \sum_{i\in\mathcal{J}_u}g_{i}^{-p}(x_{c}) \\
    &\approx \sum_{i\in\mathcal{J}_u} \phi^{-p}(x_{i} - x_{c},\sigma) \cdot \Delta x^{-p} \\
    &= \left( \frac{1}{\sqrt{2\pi}\sigma} \right)^{-p} \cdot \Delta x^{-p-1} \sum_{i\in\mathcal{J}_u} \exp{\left( p \cdot \frac{(x_{i} - x_{c})^{2}}{2\sigma^{2}} \right)} \cdot \Delta x \\
    &\approx  \left( \frac{1}{\sqrt{2\pi}\sigma} \right)^{-p} \cdot \Delta x^{-p-1} \lim_{\Delta x \to 0} \sum_{i\in\mathcal{J}_u} \exp{\left( \frac{(x_{i} - x_{c})^{2}}{\frac{2}{p}\sigma^{2}} \right)} \cdot \Delta x \\
    &\approx \left( \frac{1}{\sqrt{2\pi}\sigma} \right)^{-p} \cdot \Delta x^{-p-1} \int_{x_{c} - u}^{x_{c} + u} \exp{\left( \frac{(x - x_{c})^{2}}{\frac{2}{p}\sigma^{2}} \right)} dx \\
    &= \left( \frac{1}{\sqrt{2\pi}\sigma} \right)^{-p} \cdot \Delta x^{-p-1} \cdot 2 \int_{0}^{u} \exp{\left( \frac{x^{2}}{\frac{2}{p}\sigma^{2}} \right)} dx \\
    &= \left( \frac{1}{\sqrt{2\pi}\sigma} \right)^{-p} \cdot \Delta x^{-p-1} \cdot \frac{\sqrt{2}\sigma}{\sqrt{p}} \cdot 2 \int_{0}^{\frac{u}{\sqrt{2}\sigma} \cdot \sqrt{p}} \exp{\left( x^{2} \right)} dx \\
    &= \frac{1}{\sqrt{p}} \cdot \left( \frac{\Delta x}{\sqrt{2\pi}\sigma} \right)^{-p-1} \cdot \frac{2}{\sqrt{\pi}} \int_{0}^{\frac{u}{\sqrt{2}\sigma} \cdot \sqrt{p}} \exp{\left( x^{2} \right)} dx \\
    &= \frac{1}{\sqrt{p}} \cdot \left( \frac{\Delta x}{\sqrt{2\pi}\sigma} \right)^{-p-1} \cdot \erfi{\left( \frac{u}{\sqrt{2}\sigma} \sqrt{p} \right)} \,.
\end{split}
\end{equation}
Here the last steps come from the definition of the $\erfi$ function:
\begin{equation}
\begin{split}
    \erfi{(x)} &= -\iu\erf{(\iu x)} \\
    &= -\iu \cdot \frac{2}{\sqrt{\pi}} \int_{0}^{\iu x} e^{-v^{2}} dv \\
    &= -\iu \cdot \iu \cdot \frac{2}{\sqrt{\pi}} \int_{0}^{x} e^{v^{2}} dv \\
    &= \frac{2}{\sqrt{\pi}} \int_{0}^{x} e^{v^{2}} dv.
\end{split}
\end{equation}

\section{Proof of Proposition 2}
\label{appendSNR}
The proof follows from the application of the first derivative criterion, we see that
\begin{align}
    \frac{\partial S/N(u)}{\partial u} &=\frac{2\sqrt{2}\sigma P'\left(\sqrt{2}\sigma u\right )F \left [ P\left (\sqrt{2}\sigma u\right )F+4\sigma^2u(B-D)\right ] }{2\left [P\left (\sqrt{2}\sigma  u\right )F+4\sigma^2u(B-D) \right ] ^{3/2}} \nonumber \\
    &-\frac{ P\left (\sqrt{2}\sigma u\right )F\left [ \sqrt{2}\sigma P'\left (\sqrt{2}\sigma  u\right )F+4\sigma^2u(B-D) \right ] }{2\left [P\left (\sqrt{2}\sigma u\right )F+4\sigma^2u(B-D) \right ] ^{3/2}}  .
\end{align}
By imposing $ \frac{\partial S/N(u)}{\partial u} \biggr\rvert_{u^{\ast}_{S/N}} =0 $ and rearranging terms we obtain
\begin{equation}
\begin{split}
 &F P'\left (\sqrt{2}\sigma u^{\ast}_{S/N} \right )P\left (\sqrt{2}\sigma u^{\ast}_{S/N}\right )= \\
 &-\left [ 2\sqrt{2}\sigma u^{\ast}_{S/N}P'\left (\sqrt{2}\sigma u^{\ast}_{S/N}\right ) - P\left (\sqrt{2}\sigma u^{\ast}_{S/N}\right ) \right ]  2\sqrt{2}\sigma\left (B-D\right )  . 
\end{split}
\end{equation}

\section{Proof of Theorem 2}
\label{Appen:Theo}

The proof of Theorem 2 involves three parts. First, the elements of the Fisher matrix are approximated according to the assumption of dominant flux. The second part consists of approximating the joint Fisher information. The third part  verifies that the monotonicity conditions hold for both approximations.

\subsection{Approximations for the elements of \eqref{eq:Fisher_matrix}}
\label{sec:FI_approx}

Due to the discrete nature of the Fisher information matrix, the objective of this approach comes from the need to obtain a differentiable and continuous version of the terms in this matrix as a function of the aperture radius. Since the flux is more dominant near the central pixel, we  assume a dominant flux context for these approximations, which is equivalent to a zero-order Taylor expansion. The zero-order approximations developed in this section follow a  path similar to those presented in \citet{2013PASP..125..580M}, which though naïve, serve as a basic interpretation of how information availability depends on scenario configuration. We have the following:

\begin{equation}
\begin{split}
\mathcal{I}_{1,1}(\mathcal{J}_u) &= \sum_{i\in\mathcal{J}_u} \frac{g_{i}^{2}(x_{c})}{F \cdot g_{i}(x_{c}) + B} \\  & = \frac{1}{F} \sum_{i\in\mathcal{J}_u} \frac{g_{i}^{2}(x_{c})}{g_{i}(x_{c}) + \frac{B}{F}} \\
& \approx \frac{1}{F} \sum_{i\in\mathcal{J}_u} g_{i}(x_{c})  \approx \frac{1}{F}\cdot P(u) 
\end{split}
\label{eq:I11_ZO_FD}
\end{equation}

\begin{equation}
\begin{split}
\mathcal{I}_{1,2}(\mathcal{J}_u) = \mathcal{I}_{2,1} (\mathcal{J}_u)&= \sum_{i\in\mathcal{J}_u} \frac{g_{i}(x_{c})}{F \cdot g_{i}(x_{c}) + B} \\
& = \frac{1}{F} \sum_{i\in\mathcal{J}_u} \frac{g_{i}(x_{c})}{g_{i}(x_{c}) + \frac{B}{F}} \\
& \approx \frac{1}{F} \sum_{i\in\mathcal{J}_u} 1 \\
&= \frac{|\mathcal{J}_u|}{F} \\
&= \frac{1}{F} \cdot \frac{2u}{\Delta x} 
\end{split}
\label{eq:I12_ZO_FD}
\end{equation}

\begin{equation}
\begin{split}
\mathcal{I}_{2,2}(\mathcal{J}_u) &= \sum_{i\in\mathcal{J}_u} \frac{1}{F \cdot g_{i}(x_{c}) + B} \\
& = \frac{1}{F} \sum_{i\in\mathcal{J}_u} \frac{1}{g_{i}(x_{c}) + \frac{B}{F}} \\
 &\approx \frac{1}{F} \sum_{i\in\mathcal{J}_u} \frac{1}{g_{i}(x_{c})} \\
 &\approx \frac{1}{F}  2\pi  \left( \frac{\sigma}{\Delta x} \right)^{2}  \erfi{\left( \frac{u}{\sqrt{2}\sigma} \right)}.
\end{split}
\label{eq:I22_ZO_FD}
\end{equation}

\subsection{Approximation of the joint Fisher information}

Using the approximations for the elements of \eqref{eq:Fisher_matrix}: (\ref{eq:I11_ZO_FD}), (\ref{eq:I12_ZO_FD}) and (\ref{eq:I11_ZO_FD}) we have that

\begin{equation}
\begin{split}
\mathcal{I}_{F}(\mathcal{J}_u) & = \mathcal{I}_{1,1}(\mathcal{J}_u)- \frac{\mathcal{I}_{1,2}^{2}(\mathcal{J}_u)}{\mathcal{I}_{2,2}(\mathcal{J}_u)}\\
& \approx \frac{1}{F}\cdot P(u)  -\frac{\left ( \frac{1}{F} \cdot\frac{2u}{\Delta x}  \right )^2}{\frac{1}{F}  2\pi  \left( \frac{\sigma}{\Delta x} \right)^{2}  \erfi{\left( \frac{u}{\sqrt{2}\sigma} \right)}} \nonumber \\
&=\frac{1}{F} \left ( P\left ( u\right )  -\frac{2u^2}{\pi \sigma^2 \erfi\left( \frac{u}{\sqrt{2}\sigma}\right )}  \right )
\end{split}
\end{equation}

\begin{equation}
\begin{split}
\mathcal{I}_{B}(\mathcal{J}_u) & = \mathcal{I}_{2,2}(\mathcal{J}_u) - \frac{\mathcal{I}_{1,2}^{2}(\mathcal{J}_u)}{\mathcal{I}_{1,1}(\mathcal{J}_u)}\\
& = \frac{1}{F}  2\pi  \left( \frac{\sigma}{\Delta x} \right)^{2}  \erfi{\left( \frac{u}{\sqrt{2}\sigma} \right)}  -\frac{\left ( \frac{1}{F} \cdot \frac{2u}{\Delta x}    \right )^2}{\frac{1}{F}\cdot P(u) }\nonumber \\
& = \frac{1}{F} \left (  2\pi \left ( \frac{\sigma}{\Delta x} \right )^2\erfi\left( \frac{u}{\sqrt{2}\sigma}\right )  -\frac{4u^2}{P(u)(\Delta x)^2 }  \right ).
    \end{split}
\end{equation}

\subsection{Monotonicity of the joint Fisher information}

For the last part of the proof we see that

\begin{equation}
\lim_{u\rightarrow \infty}\frac{u^2}{\erfi\left( \frac{u}{\sqrt{2}\sigma}\right )}=0
\end{equation}
\begin{equation}
\lim_{u\rightarrow \infty}P(u)=1,
\end{equation}
which means that:
\begin{equation}
\lim_{u\rightarrow \infty}  \mathcal{I}_{F}(\mathcal{J}_u) =\frac{1}{F}.
\end{equation}
\section{Additional figure}
\label{Appen:Fig}
\begin{figure*}
\sidecaption
\begin{minipage}{12cm} 
   \includegraphics[width=\linewidth]{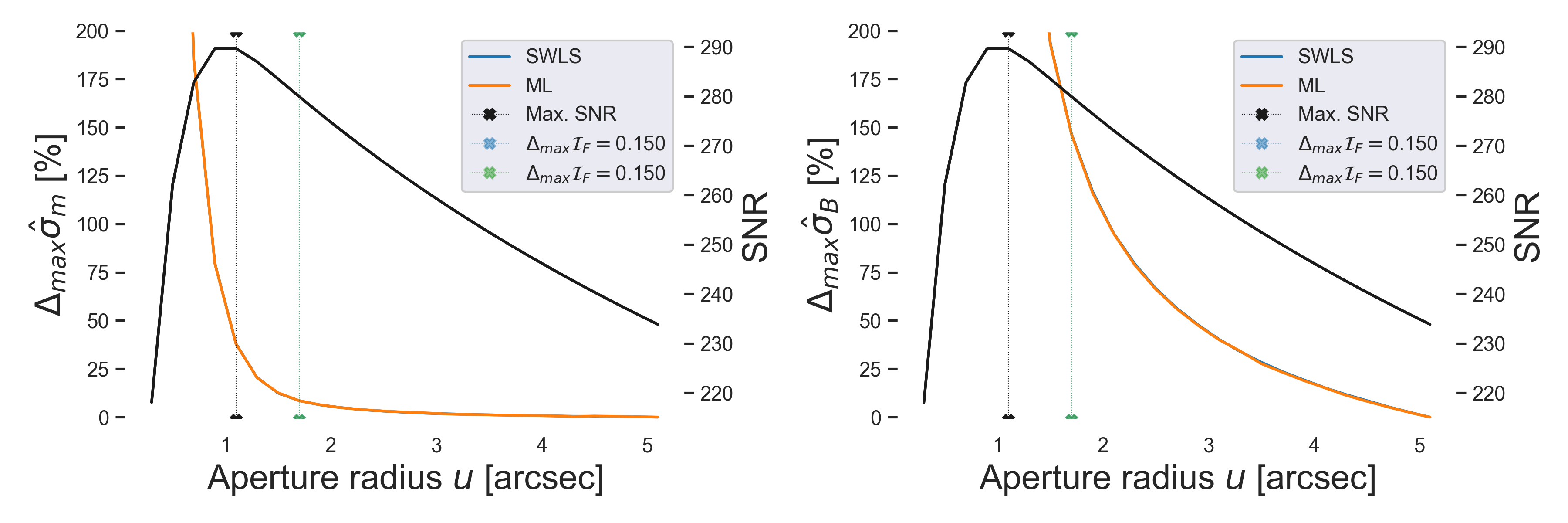}
   \centerline{(a) $F$: 100000 [photo-e$^{-}$]  -  $B$: 1625.0 [photo-e$^{-}$] FWHM: 1.0 [arcsec]}
   \includegraphics[width=\linewidth]{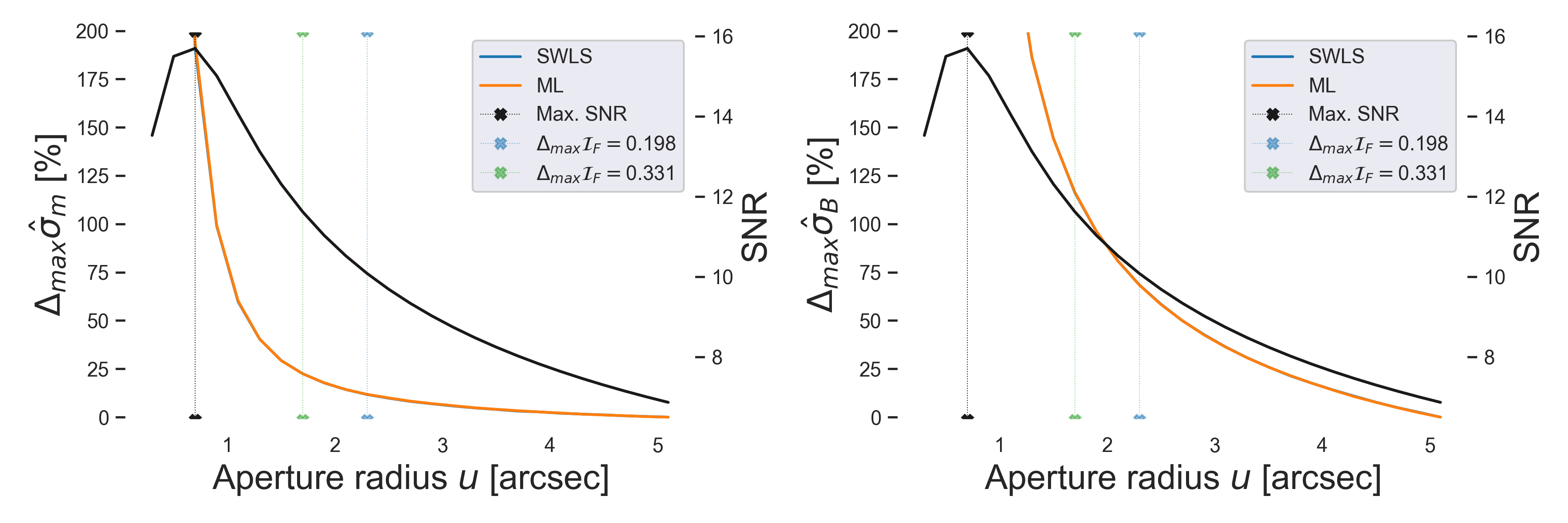}
   \centerline{(b)   $F$: 2000 [photo-e$^{-}$]  -   $B$: 1625.0 [photo-e$^{-}$] FWHM: 1.0 [arcsec]}
   \includegraphics[width=\linewidth]{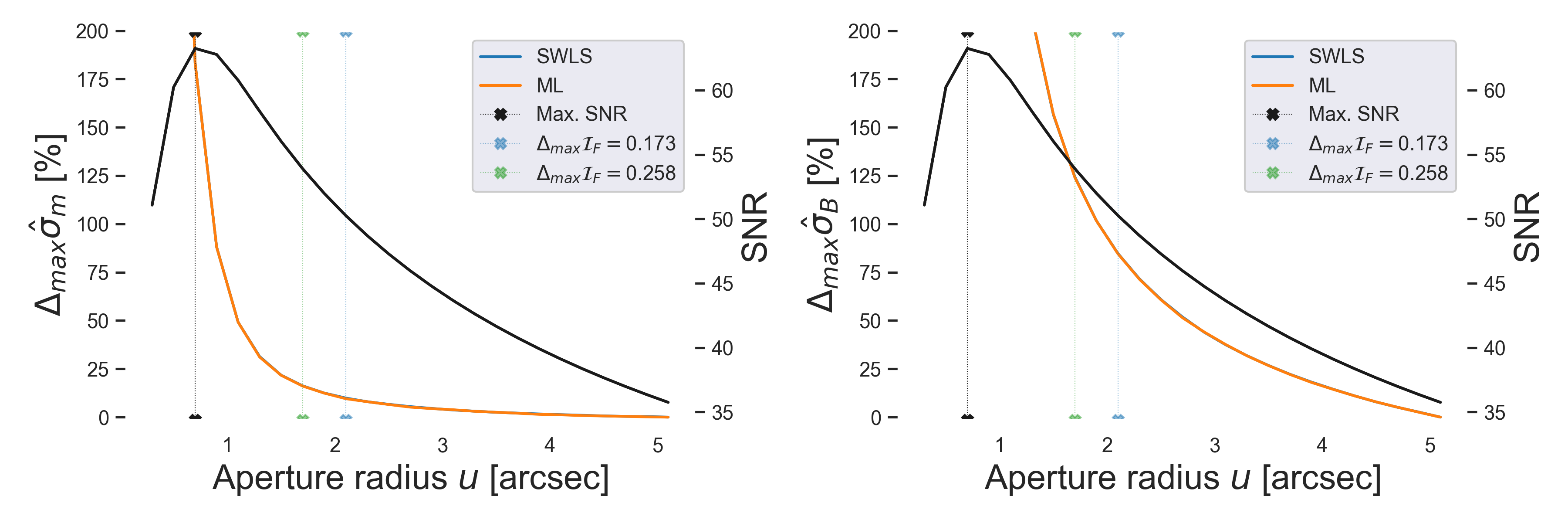}
   \centerline{(c) $F$: 8000 [photo-e$^{-}$]  -  $B$: 825.0 [photo-e$^{-}$] FWHM: 1.0 [arcsec]}
   \includegraphics[width=\linewidth]{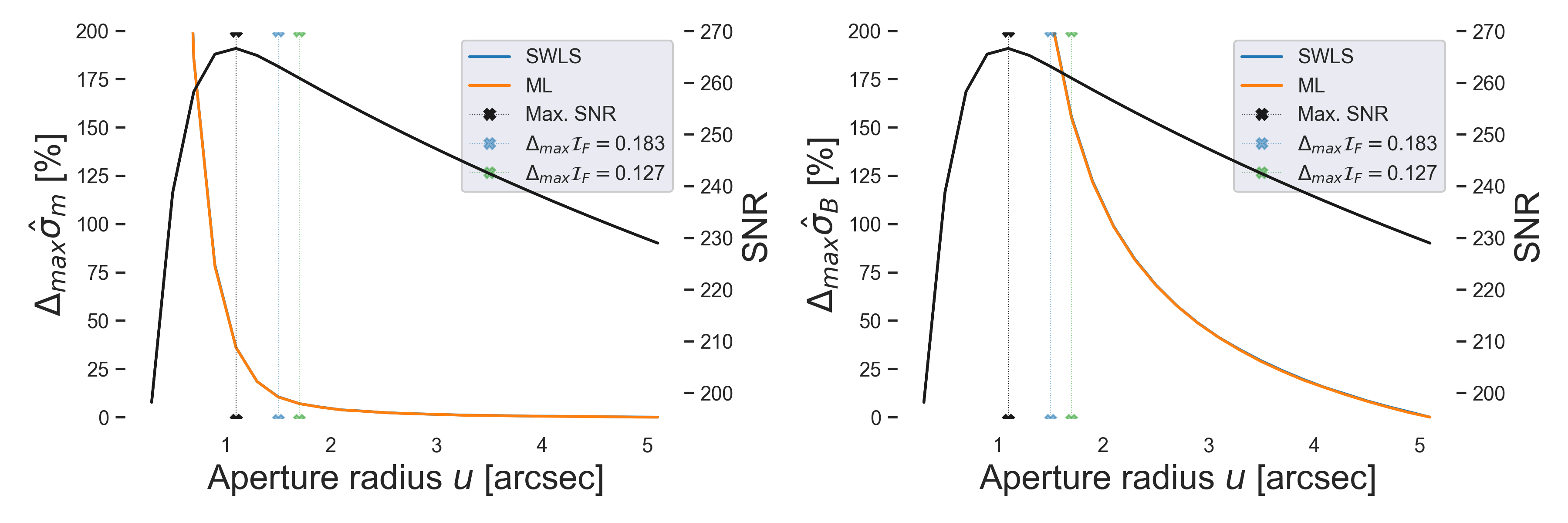}
   \centerline{(d)  $F$: 80000.0 [photo-e$^{-}$]   -   $B$: 825.0 [photo-e$^{-}$] FWHM: 1.0 [arcsec]}
\end{minipage}
\caption{Discrepancy analysis of the estimators as a function of the PSF aperture fitting radius in photometry (orange line, left ordinate on each plot, in \%). The discrepancy in the source flux (left column) is defined in Eq.~(\ref{eq:discr_f}). The discrepancy in the background (right column) is defined in Eq. (\ref{eq:discr_b}). The broad black line (right ordinate on each plot) indicates the corresponding S/N value. The vertical black line shows the aperture at which the S/N is maximized. The vertical blue line corresponds to the value of the aperture radius following the Fisher information methodology proposed in Eq.~(\ref{optiV1}), while the green vertical line corresponds to the value of the aperture radius following the approximated methodology in Eq.~(\ref{optiI}) (in some cases, the blue and green vertical lines overlap). The achieved discrepancy is observed to be close—but not necessarily equal—to the threshold value of $20\%$ due to the discrete nature of the summation. Specifically, $\Delta_{\mathrm{max}} I_F$ denotes the last value of the normalized discrepancy that remains below the selected threshold $\delta$ as the aperture radius increases, which may not exactly match $\delta$. The choice of $\delta = 20\%$ is arbitrary and is adopted here purely for illustrative purposes. Results are reported for different representative values of $F, B$ and a FWHM of 1.0~arcsec.}
\label{fig:implicit}
\end{figure*}

\end{appendix}

\end{document}